\newif\ifextended%
\extendedtrue%

\ifextended%
\documentclass[sigconf,authorversion]{acmart}
\else
\documentclass[sigconf]{acmart}
\fi
\usepackage{amsfonts}
\usepackage{bbm}
\usepackage{nicefrac}
\usepackage{cleveref}
\usepackage{algorithm}
\usepackage{tikz}
\usepackage{enumitem}

\usepackage[noend]{algorithmic}

\usepackage{hyperref}
\newcommand{\shahrzad}[1]{\textcolor{blue!60!black}{\textsc{Shahrzad:} \emph{#1}}}

\newtheorem{defin}{Definition}[section]
\newtheorem{theorem}{Theorem}[section]

\newtheorem{remark}[theorem]{Remark}

\newtheorem{lemma}[theorem]{Lemma}

\newtheorem{exm}[theorem]{Example}
\newtheorem{prob}{Problem}

\newcommand{\tmix}{\tau_{\rm mix}}
\newcommand{\trel}{\tau_{\rm rel}}
\newcommand{\cP}[1]{\mathbb P\left( #1\right)}
\newcommand{\var}{\mathbb{V}}
\newcommand{\ex}{\mathbb{E}}

\newcommand{\Ne}{\mathcal{N}}

\newcommand\algoname{{\textsc{DeMEtRIS}}}
\newcommand\algonamelayers{{\textsc{ConstructLayers}}}
\newcommand{\mypar}[1]{\medskip\noindent{\sffamily\bfseries #1.}~}

\newcommand\uniform{U}

\newcommand{\abs}[1]{\vert#1\vert}
\newcommand{\ver}{\mathsf v}
\newcommand{\V}{{\mathsf V}}
\newcommand{\g}{{\zeta}}

\newcommand{\layer}[1]{{\mathcal L}_{#1}}
\newcommand{\set}[1]
{{\mathcal G}(#1)}
\newcommand{\seg}
{\mathsf{\sigma}}
\newcommand{\cut}[2]
{\mathrm{Bnd}_{#1}\left(#2\right)}

\newcommand{\f}[1]{\mathrm{C}_{#1}}
\newcommand{\fmax}[1]{\mathrm{C}_{#1}^{\max}}
\newcommand{\assig}{\mathcal{A}}%{{ASG}}%{{\rm Assg}}
\newcommand{\assigi}[1]{{\assig}_{#1}}%{{\rm Assg}^{#1}}%{{\rm Assig}^{(#1)}}

\newcommand{\order}{\underset{V}{\prec}}
\newcommand{\Or}[1]{{\mathcal O}_{#1}}%{{\mathcal O}_{(#1)}}

\newcommand{\dg}{d}
\newcommand{\degreei}{\mathrm{D}}%{\mathrm{D}_{\seg}}
\newcommand{\degree}{\degreei}
\newcommand{\degreetalya}{\mathrm{D}}
\newcommand{\Nseg}{\Ne_{\seg}}
\newcommand{\rep}{\mathrm{R}}%{R_{\seg}}

\newcommand{\eem}{\textsc{c}}
\newcommand{\csmooth}{\eem \text{-dense}}

\newcommand{\p}{p}

\newcommand{\estimate}{\widehat{T}}

\newcommand{\gName}[1]{\texttt{#1}}
\copyrightyear{2023}
\acmYear{2023}
\setcopyright{acmcopyright}
\acmConference[WSDM '23] {Proceedings of the Sixteenth ACM International Conference on Web Search and Data Mining}{February 27--March 3, 2023}{Singapore, Singapore.}
\acmBooktitle{Proceedings of the Sixteenth ACM International Conference on Web Search and Data Mining (WSDM '23), February 27--March 3, 2023, Singapore, Singapore}
\acmPrice{15.00}
\acmISBN{978-1-4503-9407-9/23/02}
\acmDOI{10.1145/3539597.3570438}
\begin{document}

%%
%% The "title" command has an optional parameter,
%% allowing the author to define a "short title" to be used in page headers.
% \title{\algoname: counting motifs by crawling a small part of a network}

\title{\algoname: Counting (near)-Cliques by Crawling}
\subtitle{Extended version}

%%
%% The "author" command and its associated commands are used to define
%% the authors and their affiliations.
%% Of note is the shared affiliation of the first two authors, and the
%% "authornote" and "authornotemark" commands
%% used to denote shared contribution to the research.
\author{Suman K.Bera}
\authornotemark[1]
\email{sumankalyanbera@gmail.com}
\affiliation{%
  \institution{Katana Graph}
  \city{San Jose}
  \state{CA}
  \country{USA}
}

\author{Jayesh Choudhari}
\authornotemark[1]
\email{choudhari.jayesh@alumni.iitgn.ac.in}
\affiliation{%
  \institution{Cube Global Ltd.}
  \city{London}
  \country{UK}}

\author{Shahrzad Haddadan}
\authornotemark[1]
\email{shahrzad.haddadan@gmail.com}
\orcid{}

\affiliation{%
  \institution{Data Science Initiative, Brown University}
  \city{Providence}
  \state{RI}
  \country{USA}}
\affiliation{%
  \institution{Rutgers Business School}
  \streetaddress{100 Rockafeller Road, Piscataway, NJ 08854.}
  \city{Piscataway}
  \state{NJ}
  \country{USA}}

\author{Sara Ahmadian}
\email{sahmadian@google.com}
\orcid{}
\affiliation{%
  \institution{Google Research}
  \city{Mountain View}
  \state{CA}
  \country{USA}}

%%
%% By default, the full list of authors will be used in the page
%% headers. Often, this list is too long, and will overlap
%% other information printed in the page headers. This command allows
%% the author to define a more concise list
%% of authors' names for this purpose.
\renewcommand{\shortauthors}{Bera, Choudhari, Haddadan, and Ahmadian.}

\begin{abstract}
    We study the problem of approximately counting \emph{cliques} and \emph{near cliques} in a graph, where the access to the graph is only available through
crawling its vertices; thus typically seeing only a small portion of it. This model, known as the \emph{random walk} model or the \emph{neighborhood query} model has been introduced recently and captures real-life scenarios in which the entire graph is too massive to be stored as a whole or be scanned entirely and sampling vertices independently is non-trivial in it.  

We introduce \textsc{DeMEtRIS}: Dense Motif Estimation through Random Incident Sampling. This method provides a scalable algorithm for clique and near clique counting in the random walk model. We prove the correctness of our algorithm through rigorous mathematical analysis and extensive experiments. Both our theoretical results and our experiments show that \algoname\ obtains a high precision estimation by only crawling a sub-linear portion on vertices, thus we demonstrate a significant improvement over previous known results.

\end{abstract}

%%
%% The code below is generated by the tool at http://dl.acm.org/ccs.cfm.
%% Please copy and paste the code instead of the example below.
%%


%%
%% Keywords. The author(s) should pick words that accurately describe
%% the work being presented. Separate the keywords with commas.

\maketitle

\section{Introduction}
\label{sec:intro}

The subgraph counting problem (a.k.a the motif counting problem) asks for the number of copies of a small fixed subgraph (a.k.a motif) in a given large input graph. It has numerous applications in a wide variety of domains, such as, 
bioinformatics~\cite{Milo,adamcsek2006cfinder,reimann2017cliques,kose2001},
%,corander2008bayesian
cyber security~\cite{savage2014anomaly,liu2005detecting,kang2013big}, 
%,ding2019deep
and network analysis~\cite{holland1977method,jackson2012social,coleman1988social}.
Naturally, this problem has been studied extensively by both the theoretical community~\cite{eden2017approximately,eden2017sublinear,bera2017towards,kane2012cliquecounting} and the practitioners~\cite{bressanIPL,becchetti2010efficient,stefani2017triest,liu2019sampling,jowhari2005new,triangleBeraPODS,rahman104,pavan2013vldb,kane2012cliquecounting,eden2018provable}. The clique counting and the near-clique counting are arguably the most
studied variations of the subgraph counting problem~\cite{reimann2017cliques,CliqueEden2018,danisch2018listing,kane2012cliquecounting,jain2017fast,finocchi2015clique, bera2017towards}. 

Typically, most of these works' underlying model of computations assumes full access to the graph.
Even for popular restricted-access models, such as distributed and streaming models,
the underlying algorithms process information from the entire graph. Many works have also assumed the possibility of collecting independent samples from the set of vertices or edges, or querying an arbitrary edge or vertex (see \cref{sec:related}). However, in many real-world scenarios, these assumptions do not hold. The only tool 
available to the practitioners in such cases is crawling the graph through public APIs; some examples include accessing Facebook or Twitter graph via user access APIs~\cite{facebook,twitter}. In these scenarios, basic statistics about the graph, such as the number of vertices or number of edges itself, is non-trivial to estimate~\cite{haddadanicalp,HamouPeresSODA}, let alone more complicated global properties such as clique or near clique counts. Our work
takes up the challenging task of computing these global properties of the graph while looking at a tiny
fraction of the graph via crawling.

To formalize the {\em access through crawling} notion, Chierichetti et al \cite{chierichetti2016sampling}
introduced the {\em random walk access} model. This model is further explored in~\cite{dasgupta2014estimating, haddadanicalp, triangleBerakdd,ben2021sampling}. 
% In literature, this model is also known as the {\em neighborhood oracle} model.
% In this paper, we use them interchangeably.

\mypar{The random walk (access) model} In this model, an algorithm has access to the input graph $G=(V,E)$ through 
a \emph{neighborhood query oracle} $\mathcal O$.  For any $v\in V$, $\mathcal{O}(v)$ returns the set of pointers 
to the neighbors of $v$ in $G$. A typical crawling pattern for an algorithm starts from an arbitrary seed vertex $s\in V$; 
then using $\mathcal{O}(s)$ it obtains the neighborhood pointers and selects one of them uniformly at random to make
further queries to $\mathcal{O}$; thus, effectively simulating a random walk in $G$ starting at $s$.

Each query to the neighborhood oracle is expensive. The main challenge is to design algorithms with as few queries as possible.
While some basic operations such as generating a uniform random vertex sample are proven to be costly in 
this model~\cite{haddadanicalp}, there has been a few promising results in estimating simple statistics such as 
average degree~\cite{dasgupta2014estimating}, triangle counts~\cite{triangleBerakdd} etc  (see \cref{sec:related}).
In our work, we continue the algorithmic exploration by considering a more complex global statistics estimation problem: counting
copies of higher-order cliques and near-cliques. Formally, we provide a solution to the following problem.

\begin{prob}\label{prob:main}
For a fixed clique or near clique $\g$, design an algorithm in the random walk model to estimate the number
of copies of $\g$ in a given input graph while optimizing  the (sub-linear) number of queries made to the neighborhood oracle $\mathcal{O}$.
%  and given an input graph $G=(V,E)$ with $m$ edges and  $T$ many copies of $\g$, design an algorithm whose access to $G$ is through the neighborhood oracle $\mathcal{O}$, makes only sub-linear in $m$, $\epsilon^{-1}$ and $\delta^{-1}$ queries, and outputs $\estimate$ satisfying:
% \[
% \cP{\estimate \in (1\pm\epsilon) \abs{T}} \geq 1- o(1) \enspace. 
% \]
\end{prob}

\subsection{Summary of Challenges}\label{sec:challenges}

There are several fundamental challenges in tackling~\Cref{prob:main}. Some arise from the nature of the model, such as only local neighborhood access, lack of independent samples, etc. Others come from the nature of the problem: non-uniform distribution of the higher-order subgraphs in the graph leading to high variance in estimators. We summarize these challenges below.

(Challenge 1.)~
In the random walk model, sampling edges are cheaper than sampling vertices \cite{haddadanicalp}. One can run a random walk to collect \emph{almost} uniform random samples from $E$. However, the main challenge in obtaining theoretical guarantees for the usages of random walks (Markov chains) is that the samples generated by them are \emph{NOT} independent. Thus it is essential to bound the co-variance of samples. In \cref{tech:rw} we present the necessary definitions and techniques required from the theory of Markov chains. An interested reader can find the complete discussion in  \cite{levinbook}.

(Challenge 2.)~ 
In order to count the number of copies of a motif $\g$ in a graph, some methods sample edges, and for each edge $e$, they count the copies of $\g$ which contain $e$. By using a unique assignment rule which maps each copy of $\g$ uniquely to one of its edges they avoid counting duplicates. This strategy leads to high query complexity in our model due to two main factors: (1) in order to find copies of $\g$ which contain $e$, we need to query all the neighborhood of both endpoints of $e$, (2) some edges in $E$ may be  contained in no copy of $\g$ and some  in a large number, say polynomial in $\abs{V}$ of them. Thus, the variance of the estimator will be large resulting in the need for a very large (potentially more than linear) number of queries. To circumvent this issue we use a technique known as \emph{hierarchical sampling}
 which was initially proposed   for the task of clique counting \cite{CliqueCountEden2020} in Goldreich model (see related work for a brief description of Goldriech model). In \cref{tech:hir} we explain this technique, and in \cref{sec:algo} we show how we adopt this technique to work for general motifs.

(Challenge 3.) The hierarchical sampling technique in~\cite{CliqueCountEden2020} exploits the connection between the count of cliques in a graph with its arboricity (see ~\Cref{tech:arb}). However, no such 
connections are known for the count of other motifs, making it more challenging to apply hierarchical sampling techniques directly. We overcome this by proving new combinatorial results for a family of dense subgraphs that we formalize as $\csmooth$. This enables us to design provably efficient counting algorithms for  $\csmooth$ in the random walk access model.

\subsection{Summary of Our Contribution}

The main contribution of our paper is a scalable algorithm in the random walk model to count cliques and near-cliques :\algoname  - 
\textsc{D}ense \textsc{M}otif count \textsc{E}s\textsc{T}imation through \textsc{R}andom \textsc{I}ncident \textsc{S}ampling.  \algoname~ comes with theoretical guarantees, and it is the first method whose query complexity is provably  only  \emph{sub-linear} in the number of vertices and edges of the network.  In practice, \algoname\ exhibits excellent accuracy while crawling a tiny part of the graph and significantly improves  prior works.

\mypar{Theoretical Results} We present our main theoretical result in \cref{thm:intro}. To show the correctness of our methods we formalize $\csmooth$ graphs. For any $\csmooth$ motif we show the number of copies of it in a graph in terms of the graph's arboricity (\cref{lem:arboricity}). This result can serve independent interests beyond its application here. 

% \begin{itemize}[leftmargin=*]
% \item The main contribution of our paper is to provide the first scalable algorithm in the  random walk model to count cliques and near cliques: \algoname. The performance  of \algoname\ is proven through mathematical analysis and extensive experiments. 
% \item Our main theoretical contribution is presented in \cref{thm:intro}. 
% To show the correctness of our methods we formalize $\csmooth$ graphs and show that for any $\csmooth$ motif
%  the number of copies of dense motifs in a graph to its arboricity (\cref{lem:arboricity}). This result can serve independent interests beyond its application here. 
% \item We run our algorithm experimentally to show its scalability, and we compare our algorithm with other random walk based methods and show the superiority of our method. 
% \end{itemize}
\iffalse 
\begin{prob}\label{prob:main}
Given a graphlet $\g$, and a graph $G=(V,E)$ with $\abs{V}=n$, we are interested to design an algorithm whose access to $G$ is through the neighborhood oracle $\Or{G}$, makes only sub-linear in $n$, $\epsilon^{-1}$ and $\delta^{-1}$ queries, and outputs $s$ satisfying:
\[
\cP{s \in (1\pm\epsilon) \abs{\set{\g}}} \geq 1- \delta \enspace. 
\]
\end{prob}

\fi

% \paragraph{Statement of the main theorem.}

% We define a $\csmooth$ motif as follows.
\begin{defin}[$\csmooth$ motif]
\label{def:csmooth}
Consider motif $\g$ of size $k$.  We say $\g$ is  $\csmooth$ iff there is a way to divide it to induced connected sub-motifs $\g_{k-1},\g_{k-2}, \dots ,\g_2$ such that, letting $\g_k=\g$,  for each $2\leq i\leq k$ the minimum degree of $\g_i$ is at least $i-1-c$ \footnote{Alternatively we may say $\g_i$s are $(c+1)$-plexes~.}
%; for  definition~ see e.g., \cite{kplx2012,seidman1978graph}~.}. %i.e., each vertex in $\g_i$ has at most $c$ missing edges. 
\end{defin}

\iffalse 
\begin{remark} If the reader is familiar with the definition of  $k$-plexes \cite{kplx2012,seidman1978graph}, \Cref{def:csmooth} is equivalent to the following definition:

Consider motif $\g$ of size $k$.  We say $\g$ is  $\csmooth$ iff there is a way to divide it to sub-motifs $\g_{k-1},\g_{k-2}, \dots ,\g_2$ such that, letting $\g_k=\g$,  for all $2\leq i\leq k$, $\g_i$ is a $(c+1)$-plex. 
\end{remark}
\fi 

Note that a $clique$ is a (0)-dense graph, any clique missing an edge is a (1)-dense graph, an example for a (2)-dense graph with 5 vertices and 6 edges is presented in \cref{fig:ex3.1}.

%Our final result will be an algorithm called \algoname, presented in \cref{sec:algo}, whose performance is guaranteed by the following theorem: 

\begin{theorem}\label{thm:intro} Given $\epsilon$, and $\csmooth$ motif $\g$ of size $k$, and let $T$ be the number of copies of $\g$ in $G$. 
Setting $l_i$s as in \cref{lemma:nice}   
we have 
\[\mathbb{P}\left(\estimate\in (1\pm\epsilon)^k \cdot T\right)\geq 1-o(1)\enspace ,\]

where $\estimate$ is the output of \algoname. 

Furthermore, \algoname\ only queries a number of vertices:

\begin{align}
\label{qcom}
\mathcal{O}
\left(
\frac{\log n^3}{\epsilon^3(1-\epsilon)^k} \cdot \left(  \frac{mF_{\max}}{T}\trel +  \frac{m F_{\max}}{T} \cdot\sum_{i=3}^{k-1} n^{c}((c+1)\alpha)^{i-(c+1)} \right.\right.\nonumber\\ \left.\left.+~\frac{m }{T} n^{c}((c+1)\alpha)^{k-(c+1)} \right) +\frac{\log n}{\epsilon^2} \sqrt{m} \tmix 
\right)\enspace .
\end{align}

\end{theorem}
$F_{\max}$ is defined as in \cref{table:def}, and roughly speaking it is the  maximum over edges the number of copies of $\g$ assigned to it. Thus, we often have $F_{\max}\ll T$. The parameters $\trel$, $\tmix$ and $\alpha$ are defined later and known to be  $\Theta(\log n)$ in social networks \cite{leskovec2009community,benjamini2014mixing}. Thus, for small $c$, \algoname\ queries a sub-linear number of edges to output a precise approximation for $T$. 

\begin{figure}[H]
    \centering
    \includegraphics[width=0.35\textwidth]{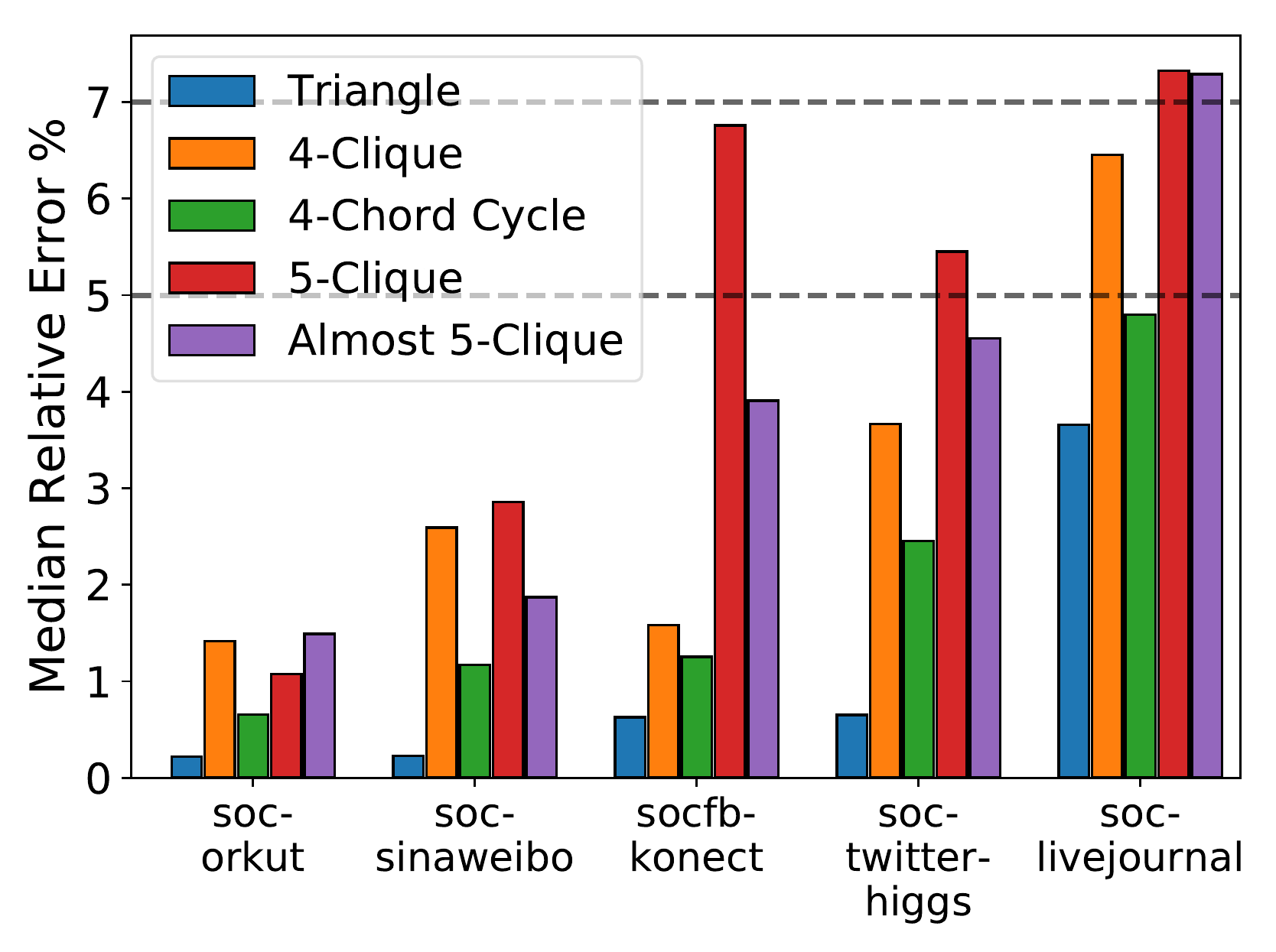}
    \caption{\small Accuracy of \algoname~ on real datasets: \gName{soc-orkut}
(3M vertices, 213M edges), \gName{soc-sinaweibo} (59M vertices,
523M edges), \gName{socfb-konect}(59M vertices, 185M edges) \gName{soc-twitter-higgs}(457K vertices, 12.5M edges), \gName{soc-livejournal}(5M vertices, 85M edges). We %run \algoname~ to 
observe at most 4\% of the edges and  repeat the experiments for
100 times. For 100 independent runs, \algoname~
achieves remarkable median relative error of less than 5\% for most of the cases.}
\label{fig:4_percent_med_err_demet_bar}
\end{figure}

\mypar{Empirical Results}
% We run our algorithm experimentally to show its scalability, and we compare our algorithm with other random walk based methods and show the superiority of our method. 
Empirically, we show that \algoname~ has excellent accuracy for a wide range of cliques and near-cliques across many large datasets (~\Cref{fig:4_percent_med_err_demet_bar}). Compared to baselines, \algoname~ has superior accuracy with cheaper query cost (~\Cref{fig:comparison_demet_srw2}). 

% \begin{figure}
%     \centering
%     \includegraphics[width=0.48\textwidth]{images/4_percent_med_rel_err_demet.pdf}
%     \caption{Median Relative Error for \algoname~ with at most $4\%$ of edges visited}
%     \label{fig:4_percent_med_err_demet}
% \end{figure}

\iffalse 

\shahrzad{put at the end}
\paragraph{\textbf{Road map}} In the rest of this introduction, we first summarize the related work, then we present the summary of our contributions and demonstrate how our work builds on prior work and extends it in various aspects. 
There are various challenges which arise from the nature of the model (local access, dependent samples), or  the nature of the problem (high order motifs are scattered in graph causing large variance of estimators). We summarize these challenges in \cref{sec:chal} and explain how we overcome them. \Cref{sec:algo} presents our algorithm and together with its theoretical guarantees, and finally \cref{sec:ex} demonstrates our experimental results.  

\fi 
\subsection{Related Work}\label{sec:related}

Prior work on counting the number of motifs are numerous (see survey \cite{seshadhri2019scalable} and references therein, e.g., \cite{
%bressanIPL,
becchetti2010efficient,stefani2017triest,liu2019sampling,jowhari2005new,triangleBeraPODS,pavan2013vldb,kane2012cliquecounting}).
A large subset of these works concentrates on counting cliques %(exploiting nice theoretical findings)
\cite{reimann2017cliques,CliqueEden2018,danisch2018listing,kane2012cliquecounting,jain2017fast,finocchi2015clique, bera2017towards,seidman1978graph}; in practice  counting  quasi-cliques is equally important (a motif in which all vertices are connected to all other by only $c$ vertices) and  has been the focus of study of some works \cite{jiang2009mining,pei2005mining,mitzenmacher2015scalable,kplx2012}.
Among all of these works we focus on those whose objective is to query \emph{a sub-linear} number of edges or vertices in the graph; thus they do not require to query  all of $G$. There are two main models which fall into this category:

\paragraph{Goldriech model \cite{goldreichbook}} In this model the algorithm has access to the graph by uniformly at random sampling $v\in V$ and performing one of the following operations (1) degree queries: at each point of time the algorithm query the degree of $v$; $d(v)$, (2) Neighborhood query: for  $v$ and $i\leq d(v)$, the algorithm obtains the $i$th neighbor of $v$ (3) the algorithm can query if two u.a.r samples $v$ and $u$ are adjacent. 

In this model 
%has been studied a lot by theoretical computer scientists and tight bounds have been obtained. 
Eden et al obtained an algorithm for counting triangles \cite{eden2017approximately} and cliques  \cite{CliqueEden2018}. Later they showed a more efficient algorithm for clique counting, in sparse graphs (exploiting the graph' arboricity ordering)  \cite{CliqueCountEden2020}. Other works concentrated on
approximating the degree distribution \cite{eden2017sublinear,eden2018provable}. 
\paragraph{Edge query model} In this model, in addition to the queries allowed by the Goldreich model, the algorithm is allowed to also use (4) sample a uniformly at random \emph{edge} from the graph. 

In this model, the problem of motif counting was studied by Assadi et al \cite{Assadi2019ASS} who proposed an algorithm based on a decomposition of the motif the stars and odd cycles. This work is designed on counting arbitrary motifs with general structure and does not concentrate on dense motifs.  
And \cite{akbarpoorstar} showed how to count stars. 

Other works studied similar access models {\cite{tvetek2022edge,Eden2018OnSE}}.

\smallskip

%Some other works have concentrated towards establishing a connection between the two models, asking questions such as: what is the computational cost of generating a u.a.r. edge in Goldreich framework \cite{hashedbased2021sampling,eden_et_al:OASIcs:2018:8300}.

While the above works have set up a beautiful theoretical framework; leading to the development of elegant results, they are not directly applicable in the random walk model.

\paragraph{Random walk model}
The random walk model was initially introduced and studied by \cite{chierichetti2016sampling}. Some works studied preliminary problems such as finding global parameters of a graph e.g., number of vertices or edges \cite{HamouPeresSODA,haddadanicalp,tvetek2022edge,cckatzir}, clustering coefficient \cite{cckatzir} or generating  uniformly at random vertices \cite{haddadanicalp,ben2021sampling}.

For counting three vertex motifs,  Rahman et al~\cite{Rahman:2014} proposed a huristic, and 
Bera et al 
\cite{triangleBerakdd} showed a rigorous method  for  approximating the number of triangles.

The problem of \emph{motif counting in the random walk model  beyond three nodes} has been explored in a few works, but yet it does not have an efficient solution. The first attempt for estimation of high-order motifs using random walks was the work of \cite{gius}, a.k.a. GUISE, who proposed   a Metropolis-Hastings-based algorithm. The performance of GUISE  suffers from  high rate of rejecting samples which makes the algorithm non-scalable in massive graphs \cite{ribeiro2010estimating,gjoka2011practical}.  
Wang et al \cite{wang2014efficiently} proposed an algorithm based on running a Markov chain on the  space of all motifs of size $k-1$ in-order to, approximately, sample and count motifs of size $k$. A major improvement was the work of
 Chen et al 
\cite{Chen2016} who  designed an algorithm based on running a random walk on motifs of size $d\ll k$ in-order to, approximate, the frequencies of motifs of size $k$. In particular letting $\g^{(1)},\g^{(2)}\dots \g^{(m)}$ be all the distinct (up to isomorphism) $k$-node motifs, and $C^{(i)}$ the count of the $\g^{(i)}$ they estimate $\rho_i=\frac{C^{(i)}}{\sum_{j=1}^m C^{(j)}}$~.  Chen et al 
\cite{Chen2016} bound the query complexity of estimating $\rho_i$s by 
\[O\left(
\frac{W}{\Lambda}\tmix \frac{1}{\epsilon^2}\log \left(\frac{1}{\delta}\right)\right);  ~ ~W=\max_X \pi(X)^{-1},
\]
where $X$ denotes a state of a so call \emph{extended} Markov chain and $\pi(X)$ refers to the probability density of $X$ at stationary distribution, and $\Lambda$ is a function of $C^{(i)}$s. Note that $\frac{W}{\Lambda}$ can be exponential in terms of number of edges of the original graph. Furthermore because the above bound is proven for estimation of $\rho_i$,  the complexity of estimating $C^{(i)}$ requires a multiplicative factor of $\sum_{j=1}^m C^{(j)}$~.

In this paper, we provide the \emph{first} algorithm in
the random walk model to count cliques and near-cliques whose query complexity is \emph{sub-linear} in the number of network's edges.  
The query  complexity of our
method, shown in \cref{qcom}, may be compared to the query complexity of Bera et al's \emph{sub-linear} triangle counting algorithm 
which bounded the query complexity  in terms of the mixing time of the chain $\tmix$, total number of triangles $T$, arboricity of the graph $\alpha$ and total number of edges $m$ as:
\[O\left(\frac{\log n}{\epsilon^2}\cdot \left(\frac{m \tmix\cdot T_{\max}}{T}+\frac{m\alpha}{T}+\sqrt{m}\tmix\right)\right) ~.\] 

In terms of techniques, our work is most similar to, and complements  the work on clique counting in the Goldreich model (\cite{CliqueCountEden2020}).
%, and triangle counting in random walk model \cite{triangleBerakdd}.
Similar to their work, we use an assignment rule and hierarchical sampling to cleverly bound the variance of our estimator.
Our work builds on these works and extends them  to  resolve various issues highlighted  in \cref{sec:challenges}.

\section{Notations and Preliminaries}

Consider a graph $G=(V,E)$ 
with $\abs{V}=n$ and $\abs{E}=m$. For any vertex $v\in V$ we denote the neighborhood of $v$ in $G$ by $\Ne(v)$ and its degree by $\dg(v)$. We further generalize these definitions for a subset $S\subseteq V$ as: $\Ne(S)=\bigcup_{v\in S}\Ne(v)$ and $\dg(S)=\vert \Ne(S)\vert $. %We use $\Ne_{\g}(\omega)$ to denote the neighborhood of a vertex $\omega$ in motif $\g$.
We define the \emph{neighborhood oracle} $\Or{G}:V\rightarrow 2^{V}$ be an oracle which given any vertex $v\in V$ returns a set of pointers to $v$'s neighborhood, i.e., $\Ne(v)\subseteq V$. % We use $\delta(G)$ to denote the minimum degree in a given graph $G$.
 Throughout the paper, we fix some lexicographic order $\order$ on $V$ for breaking ties.    %Note that through $\Or{G}$, an algorithm can start from a \emph{seed} vertex $v$, query $\Ne(v)=\Or{G}(v)$, pick $u$ w.r.t. an arbitrary distribution from $\Ne(v)$,  query $u$, and call $\Or{G}(u)$ and by repeating these steps the algorithm \emph{crawls} all of the graph. 
%Thus we are interested in an approximation for $\abs{\set{\g}}$.

\iffalse 
\begin{prob}\label{prob:main}
Given a graphlet $\g$, and a graph $G=(V,E)$ with $\abs{V}=n$, we are interested to design an algorithm whose access to $G$ is through the neighborhood oracle $\Or{G}$, makes only sub-linear in $n$, $\epsilon^{-1}$ and $\delta^{-1}$ queries, and outputs $s$ satisfying:
\[
\cP{s \in (1\pm\epsilon) \abs{\set{\g}}} \geq 1- \delta \enspace. 
\]
\end{prob}
\fi 

Our focus in this paper is counting $\csmooth$ motifs. We fix the notation and use $\g$ to denote the graphlet (motif) of interest with $k$ vertices ($k$ being a constant). An induced subgraph $g$ is a copy of $\g$ if $g$ and $\g$ are isomorphic, denoted mathematically $g\equiv \g$. We denote the set of all copies of $\g$ in $G$ by ${\mathcal G}_G(\g)$, simplified to $\set{\g}$ when $G$ is clear from the context. Throughout the paper, to refer to motifs, we use Greek letters like $\g$, $\sigma$, etc, and we refer to motifs' copies in $G$ using lowercase English letters $g$, $h$, etc. For example let $\g$ be a triangle, we may write $g\in \set{\g}$, which means vertices of $g$ forms a triangle in $G$. 

Consider two induced subgraphs $g$ and $h$ in $G$, we say $g$ is a sub-motif of $h$ if $g$ is an induced subgraph of $h$, denoted by  $g\subseteq h$; we use the same notation and terminology for graphlets. %  $\g$ and $\gamma$.
   We may add or remove a vertex from or to a subgraph (of $G$) to obtain a new subgraph. To denote these operations, we use $+$ and $\setminus$, e.g., $g'=g\setminus v$ means $g'$ is an induced subgraph of $g$ which is obtained by removing $v$ from $g$, and equivalently $g = g' + v$ which is obtained by adding $v$ to $g$ along with edges between $v$ and $g$.

%There are several challenges that arise for the design of algorithm, in the next section we describe these challenges and show how we address them using various techniques. We show that our algorithm is fast and beats the state of the art for counting dense motifs. 
We would like to be able to uniquely assign each $g\in \set{\g}$ to a sequence of its sub-motifs, through an \emph{assignment rule}.
The following definition for segmentation of a graphlet $\g$ is key to our assignment rule:

%Let $g$ be an arbitrary motif (graphlet) that is $g_k$ or an induced sub-graph of it.We say $g_r$ is a sub-motif of $g$ if $g_r$ is an induced sub-graph of $g$.We denote the number of $g$'s copies in $G$ by $\cnt{G}{g}$. 

\begin{defin}[Segmentation \& \eem-dense segmentation]\label{Def:segment}
Given a graphlet $\g$ with $k$ vertices, let $ \seg_2,\seg_3,\dots , \seg_{k-1}, \seg_k$ be a sequence of submotifs of $\g$ where each $\seg_i$ has $i$ vertices, $\seg_i\subseteq\seg_{i+1}$ for $i=2\dots k-1$, and $\seg_i$ is connected.
We call such a sequence a \emph{segmentation} of $\g$ and we denote it by $\seg(\g)=\langle \seg_2,\seg_3,\dots , \seg_{k-1}\rangle $, simplified to $\seg$ when $\g $ is clear from the context.
%We use $\omega_i$ to denote $\seg_{i}\setminus \seg_{i-1}$ which serves as label to refer to it. Given $\seg$, we define $\cut{\seg}{i}$ to be the vertex set in the $\seg_i$  connecting it to $\seg_{i+1}$, i.e., 
%\[  \cut{\seg }{i} \doteq  \{v \in \seg_{i} : \omega_{i+1}\in\Ne_{\g}(v)  \}= \Ne_{\g}(\omega_{i+1})\cap \seg_i\enspace.\]

A segmentation $\seg$ is called {\csmooth} if each vertex of $\seg_i$ has degree at least $i - 1 - c$. %, i.e., each vertex in $\seg_i$ is not connected to at most $\eem$ vertices in $\seg_i$.
\end{defin}

\begin{figure}[h]
\begin{tikzpicture}
\filldraw[black] (1,1.2) circle (1pt);
\filldraw[black] (0.5,1) circle (1pt);
\filldraw[black] (1,1.2) -- (0.5,1);
\node[label=below:$\seg_2$] at (.75,0.2) {};

\filldraw[black](2,1.2) circle (1pt);
\filldraw[black] (1.5,1) circle (1pt);
\filldraw[black] (2.5,1) circle (1pt);
\filldraw[black] (2,1.2) -- (1.5,1);
\filldraw[black] (2,1.2) -- (2.5,1);
\node[label=below:$\seg_3$] at (2,0.2) {};

\filldraw[black](3.5,1.2) circle (1pt);
\filldraw[black](3.5,0.8) circle (1pt);
\filldraw[black] (3,1) circle (1pt);
\filldraw[black] (4,1) circle (1pt);
\filldraw[black] (3.5,1.2) -- (3,1);
\filldraw[black] (3.5,1.2) -- (4,1);
\filldraw[black] (3.5,0.8) -- (3,1);
\filldraw[black] (3.5,0.8) -- (4,1);
\node[label=below:$\seg_4$] at (3.5,0.2) {};

\filldraw[black] (5,0.5) circle (1pt);
\filldraw[black] (4.5,1) circle (1pt);
\filldraw[black] (5,1.2) circle (1pt);
%\node[label=below:$\omega_5$] at (5,0.5) {};
\filldraw[black] (5,0.8) circle (1pt);
%\node[label=above:$\omega_4$] at (5,0.7) {};
\filldraw[black] (5.5,1) circle (1pt);
%\node[label=right:$\omega_3$] at (6,1) {};
\filldraw[black] (5,1.2) -- (4.5,1);
\filldraw[black] (5,1.2) -- (5.5,1);
\filldraw[black] (5,0.8) -- (4.5,1);
\filldraw[black] (5,0.8) -- (5.5,1);
\filldraw[black] (5,0.8) -- (5,0.5);
\filldraw[black] (4.5,1) -- (5,0.5);
\filldraw[black] (5.5,1) -- (5,0.5);
%\node[label=below:$\omega_5$] at (6.6,0.2) {};
\node[label=below:$\seg_5 \text{ = }\g $] at (5,0.2) {};
%\node[label=below:$=$] at (5.2,0.2) {};
%\node[label=below:$\g$] at (5.4,0.2) {};

% \filldraw[black] (6.6,0.2) circle (1pt);
% \filldraw[black] (6,1) circle (1pt);
% \filldraw[black] (6.6,1.4) circle (1pt);
% \node[label=below:$\omega_5$] at (6.6,0.2) {};
% \filldraw[black] (6.6,0.8) circle (1pt);
% \node[label=above:$\omega_4$] at (6.6,0.7) {};
% \filldraw[black] (7.2,1) circle (1pt);
% \node[label=right:$\omega_3$] at (7.2,1) {};
% \filldraw[black] (6.6,1.4) -- (6,1);
% \filldraw[black] (6.6,1.4) -- (7.2,1);
% \filldraw[black] (6.6,0.8) -- (6,1);
% \filldraw[black] (6.6,0.8) -- (7.2,1);
% \filldraw[black] (6.6,0.8) -- (6.6,0.2);
% \filldraw[black] (6,1) -- (6.6,0.2);
% \filldraw[black] (7.2,1) -- (6.6,0.2);

\end{tikzpicture}
    \caption{\small Example of a segmentation.}
    \label{fig:ex3.1}
\end{figure}
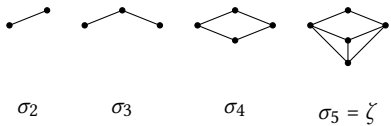

\begin{exm}
 \Cref{fig:ex3.1} shows a segmentation of a diamond shape motif of size five. This segmentation is $2$-dense as each newly added vertex is not connected to at most $2$ vertices. %For each $i=2,3,4$, the set $ \cut{\seg }{i}$ is shown in red. 
\end{exm}

To estimate $\abs{\set{\g}}$, a hierarchical sampling procedure extends any sampled  copy of $\seg_i$ in $G$ to  a copy of $\seg_{i+1}$ in $G$  by adding one more vertex from the neighborhood of $\seg_i$ in $G$ (defined in \cref{def:degree}).
%to it from  $\cut{\seg}{i}$'s neighborhood. Thus, it  is crucial to obtain a segmentation $\seg$ satisfying:
\iffalse 
\begin{equation}\label{eq:cutsets}
    \forall \  i\in \{2,3 \dots ,k-1\}, \ \abs{\cut{\seg}{i}} > 0.
\end{equation}
\fi 

%Note that since $\g$ is connected, a segmentation in which \cref{eq:cutsets} holds always exists. 

%\shahrzad{change?} \shahrzad{Is there any connection to c-closed graphs?} We call a segmentation \textbf{ $\mathbf c$-smooth} if we have $\forall i~~ \cut{i}{\seg(\g)}\geq i+1-c$.

 %their definition was then used for triangle or clique counting. 
Given $\g$, a segmentation of it $\seg$ and $ i\in \{2,3,\dots , k-1\}$,  consider the set of all copies of   $\seg_i$  in $G$, i.e., $\set{\seg_i}$. For any $g\in \set{\seg_i}$, we define its neighborhood and degree with respect to $\seg$ as follows:

\begin{defin}[Neighborhood/Degree of a motif w.r.t. a segmentation]\label{def:degree} 
Given $\g$, its \csmooth~segmentation $\seg(\g)$, and a subgraph $g\in \set{\seg_i}$, let the \emph{representative} of $g$, denoted by $\rep(g)$, be a set of $\eem + 1$ vertices in $g$ with smallest neighborhood (using $\order$ to break ties in favor of the lexicographically smallest subset). We define the {\emph neighborhood with respect to $\seg_i$} of a subgraph $g$, denoted by $\Nseg(g)$, to be the neighborhood of the subgraph $\rep(g)$. More formally, 
\[\Nseg(g)\doteq
\Ne(\rep(g)) \text{ , where } \rep(g) \doteq \underset{S\subset g: \abs{S}= \eem + 1}{{\rm argmin}} \vert \Ne(S)\vert\enspace. \]  
%The \emph{degree with respect to $\seg_i$} of a subgraph $g$ is defined as $\degreei{i}{\seg}(g)\doteq \abs{\Nseg(g)}$.
The \emph{degree with respect to $\seg_i$} of a subgraph $g$ is defined as $\degreei(g)\doteq \abs{\Nseg(g)}$.
\end{defin}

Note that because the segmentation is $\csmooth$, we are always able to obtain a vertex $v$ in $\Nseg(g_{i-1})$ to obtain $g_i$.
 We are now ready to present our hierarchical  assignment rule which assigns a subgraph $g\in \set{\g}$ to a sequence of its sub-motifs $g_{k-1},g_{k-2},\dots g_2$ so that each $g_i\equiv \seg_i$. %In principle the assignment rule involves matching for $i=3,\dots k$,   $\omega_i$s  which  label vertices of $\g$ \emph{uniquely} to  an ordering of vertices $\ver_3,\ver_4,\dots , \ver_k$ which are of vertices  in $G$.  

\iffalse
 We are now ready to present our hierarchical  assignment rule which 
 %determines how an
 %instance $\set{\seg_i}$ 
 %occurrence of a sub-motif of size $i$ 
% should be assigned to a  sub-motif in $\set{\seg_i} $
 %of size $i-1$. Applying this assignment rule consecutively on $g$ results in a sequence of its subgraphs $g_{k-1},g_{k-2},\dots g_2$ such that each $g_i\equiv \seg_i$. Moreover, using this sequence, we can uniquely determine $\ver_3,\ver_4,\dots , \ver_k$  % In principle the assignment rule involves 
 for $i=3,\dots k$,  
 \emph{uniquely}  orders  vertices of $g\in\set{\g}$ to $\ver_3,\ver_4,\dots \ver_k$ so that each $\ver_i$  matches $\omega_i$ in $\g$. Using this ordering we can generate $\g$ by consecutively building $g_i\in \set{\seg_i}$s.  
\fi

%we define $\assigi{i}:\set{\seg_{i+1}}\rightarrow \set{\seg_i}$ as follows:for any $g_{i+1}\in \set{\seg_{i+1}}$ let $S$ be a subset of $g_{i+1}$'s vertices satisfying $u\in S$ iff $g_{i+1}\setminus u\equiv \seg_i $. Let $v$ be the minimum  vertex in $S$ w.r.t. $\order$. We define \[\assigi{i}(g)\doteq g\setminus v\enspace. \]

\begin{defin}[Assignment function]\label{Def:assign}
Assume a motif $\g$ and its segmentation $\seg=\langle \seg_2,\seg_3,\dots , \seg_{k-1}\rangle$. For $2\leq i\leq k-1$ an \emph{assignment function} $\assigi{i}: \set{\seg_{i+1}} \rightarrow \set{\seg_{i}}$ satisfies the following conditions  for every $g \in\set{\seg_{i+1}}$: 
\begin{enumerate}
    \item $\assigi{i}(g) = g \setminus \ver \in \set{\seg_i}$
    \item $\ver$ is the minimum vertex w.r.t $\order$ that satisfies (1). 
\end{enumerate}
\end{defin}

\begin{defin}[Count function]\label{Def:count}
We define the \emph{count} function based on the inverse of a given assignment function (for a graph). More precisely, for each $i = 2, 3, \cdots, k$, let $\f{i}: \set{\seg_{i}} \rightarrow \mathbb{R}$ to be defined as $\f{i}(g) = |\assigi{i}^{-1}(g)|$, i.e., the number of motifs in $\set{\seg_{i+1}}$ assigned to $g$. 

Let $\fmax{i}$ be defined as the maximum value that this function takes given the assignment function, i.e., $\fmax{i} = \max_{g\in \set{\seg_{i}}} \f{i}(g)$.
\end{defin}

\subsection{Main Tools and Techniques}\label{sec:chal}

\subsubsection{Random walk sampling and an unbiased estimator.}\label{tech:rw} A simple random walk on $G$ starts \emph{crawling} its vertices at a \emph{seed vertex}, at any point of time, being at $v\in G$, it obtains $\Ne(v)$ by calling $\Or{G}(v)$, samples  $u\sim \Ne(v)$ uniformly at random, and repeats the same steps for $u$.
Let $\pi_t$ be the distribution of the random walk after taking $t$ steps. 
It is known that, under some simple conditions known as ergodicity, a simple random walk converges to a \emph{stationary distribution} $\pi$ on $V$ defined as $\pi(v)\doteq\frac{d(v)}{2\abs{E}}$ regardless of the starting distribution i.e., $\lim_{t\rightarrow \infty}\pi_t=\pi$. Let $ \rm {d_{tv}}$ denote the total variation distance,  
the \emph{mixing time} $\tmix$ of the chain is the minimum $\tau$ satisfying $ \forall t\geq \tau ;  \rm {d_{tv}}(\pi_t, \pi)\leq 1/4$ regardless of the starting point. %Furthermore it is known that for any arbitrary $\epsilon\leq 1/4$ we have $ \forall t\geq \tmix \log(\epsilon^{-1}) ;  \rm {d_{tv}}(\pi_t, \pi)\leq \epsilon$~.
Note that since $\pi(v)\propto d(v)$ it is equivalent to the uniform distribution on $E$ which we denote by $U(e)=\nicefrac{1}{\abs{E}}$, and for simplicity just $U(E)$. 
\iffalse 
\subsubsection{Unique assignment} An \emph{assignment rule} is a function $\assig:\set{\g}\rightarrow E$ which assigns to each $g\in\set{\g}$ one of its edges. Based on it, we define a real valued ${f}_{\assig}:E\rightarrow \mathbb{R}$ 
on  $G$'s edges such that for any $e\in E$, $f_{\assig}(e)$ is equal to the number of $g_k\in\set{\g} $ which satisfy $e=\assig(g_k)$. i.e., ${f}_{\assig}(e)$ is the size of inverse image of $\assig$.
Thus, $\sum_{e\in E}f_{\assig}(e)=\abs{\set{\g}}$. 

Letting $\bar{f}=\mathbb{E}_{e\sim U(E)}[f_{\assig}(e)]$ and $m=\abs{E}$, we have $m\cdot f_{\assig}$ is an \emph{unbiased estimator} for $\abs{\set{\g}}$.
Thus, by running the random walk for, say, $r$ steps and using probability concentration bounds we will be able to bound the deviation from mean as $r$ increases. 

\fi

Assume having a function $f:V\rightarrow \mathbb{R}$ and we have obtained $r$ edges $\{e_i\}_{i=1}^r$ from $E$ by running a random walk on it. We are interested to estimate $\mathbb{E}(f)$ by its empirical average  $\sum_{i=1}^r f(e_i)/r$.
Note that since the samples are not independent, the classic i.i.d bounds are not applicable.
Thus, we use  concentration bound which hold in MCMC settings e.g., \cite{Paulin2012ConcentrationIF,cousins2020making},
%chung2012chernoff,Jiang2018BernsteinsIF}
these bounds depend on the chain's mixing time $\tmix$ (defined before) or relaxation $\trel$ defined as follows. 

\begin{defin}[relaxation time and the second largest absolute eigen-value]
Let $T$ be the transition matrix of a Markov chain,  and $\lambda$ its second largest eigen-value, the relaxation time of the Markov chain is denoted by $\trel$ and it is defined as: $\trel\doteq \frac{1}{1-\lambda}$ .   
\end{defin}
 It is well known that mixing and relaxation times are related as $ (\trel -1)\log(1/8)\leq \tmix \leq \trel \log \frac{1}{4\min_{v\in V}\pi(v)}$ ; See e.g., \cite{levinbook}.
 
The following result hold for the %variance and 
co-variance of Markov chain generated samples; for full discussion see e.g., \cite{levinbook}: 

\iffalse 
\begin{theorem}[Variance of average  in MCMC dependent data \cite{Paulin2012ConcentrationIF}]\label{thm:traceVariance}
Let $X_1, X_2,\dots , X_n$ be $n$ consecutive steps of a Markov chain, and $f$ a real valued function on Markov chain's state space. We have:
\begin{equation*}
    \mathbb{V}_{\pi} [\frac{f(X_1)+f(X_2)+\dots + f(X_n)}{n}] \leq {\frac{2\trel}{n} \mathbb{V}_{\pi}[f]},
\end{equation*}
where $\pi$ is the stationary distribution of the chain. 
%and $\gamma$ is the spectral gap, known to follow: $\gamma^{-1}\leq \tmix$.
\end{theorem}
\fi

\begin{theorem}[Co-variance of MCMC samples]
\label{lem:cov}
 Let $\lambda$ be the second largest eigen-value of a Markov chain. Let $X_1,X_2,\dots X_n$ be $n$ consecutive steps of the Markov chain, for any $i,j$
 the covariance of  $f(X_i)$ and $f(X_j)$ follows: 
 \[\abs{j-i}\geq k \implies \mathbb{C}(f(X_j),f(X_i))\leq \lambda^{k}\mathbb{V}(f)~,\]
 where $\mathbb{V}(f)$ is variance of $f$ and $\mathbb{C}$ denotes the covariance of two random variables 

\end{theorem}

\iffalse 

Letting  $\V=\var_{e\sim \uniform}[f_{\assig}(e)]$ and using  \cref{thm:traceVariance}  together with the Chebyshev's bound, a straightforward solution is to crawl the graphs in $\Theta(\trel \V)$ steps to approximate $\bar{f}$. The problem with this rather easy solution is $\V$ is generally large. \shahrzad{maybe an example}. 
\fi 

\subsubsection{Hierarchical sampling and    motif segmentation} \label{tech:hir}
%To circumvent this issue we use \emph{

 An \emph{assignment rule} is a function $\assig:\set{\g}\rightarrow E$ which assigns to each $g\in\set{\g}$ one of its edges. 

The key idea behind hierarchical sampling is to design a \emph{hierarchical assignment rule} $\assig$ which is defined to be a composition of $k-2$ functions $\assigi{k-1}$,
$\assigi{k-2},
\dots,\assigi{2}$, i.e., $\assig=\assigi{2}\circ\assigi{3}\circ
\dots\circ\assigi{k-1}$ (as  in \cref{Def:assign}). 
In hierarchical sampling we  initially sample a collection of edges.  
%Our algorithm crawls the graph through a random walk and collect edges, the collected 
The edges in the collection are then extended to graphlets with $3$, $4$ and eventually $k$  vertices, and by checking  $g_{i-1}=\assigi{i-1}(g_i)$ we ensure each intermediate subgraph $g'$ eventually gets counted once and only if  $g'\subseteq g\in \set{\g}$.
In simple words, by using hierarchical sampling, as we inductively add vertices to the graphlets,  we \emph{trim} the sample space thus we manage to reduce the effective required sample size. 

% Our hierarchical assignment rule is presented in \cref{Def:assign} and it is based on what we define as \emph{a  segmentation of $\g$} (see \cref{Def:segment}), we detail the hierarchical sampling procedure and show how it is designed to make feasible estimating $\abs{\set{\g}}$ through hierarchical sampling.

%The range of each $\assigi{i}$, $S_i$, is the domain of $\assigi{i-1}$ ($S_k=\set$ and $S_2=E$), and through them each $g\in {\mathcal G}_\g(G)$ is mapped inductively to its induced sub-motifs  $e=g_2\subseteq g_3\subseteq g_4\subseteq \dots g_{k-1}\subseteq g=g_k$ where $g_i\in S_i$ and  $g_{i-1}=\assigi{i-1}(g_i)$ for $3\leq i\leq k$.   

%For each $g_i\in S_i$, let $f_{\assigi{i}}(g)$ be the number of $g_{i+1}\in S_{i+1}$ which are assigned to it. 

%Our proposed algorithm preforms a \emph{hierarchical sampling procedure}. It initially crawls the network though a random walk  to sample a subset of edges. It then sample each edge w.r.t. probability distribution $\pi_2$ and  extends these them to induced sub-motifs of size $3$, the new motifs are sampled w.r.t. distribution $\pi_3$ and extended to motifs of size  $4$ and this procedure continues until we have a motif with $k$ vertices. At each step we check  $g_{i-1}=\assigi{i-1}(g_i)$ holds. Defining $\V_i=\var_{g\sim \pi_i(S_i)}[f_{\assigi{i}}(g)]$, the number of samples we need is now dominated by $\sum_{i=2}^k{\V_i}$. In simple words using \emph{hierarchical sampling} we \emph{trim} the sample space gradually so we decrease the effective variance. 

\subsubsection{Graph's Arboricity}
\label{tech:arb}
The \emph{arboricity} of an undirected graph is the minimum number of forests into which its edges can be partitioned. Arboricity is tightly connected to the density of a graph and it is known to be small for real-life networks. More precisely, for a given clique $c_k$, let the degree of $c_k$, denoted by $\degreetalya(c_k)$, be the the degree of minimal-degree vertex (breaking ties by $\order$). \citet{CliqueCountEden2020} has established a connection between the total degree of cliques in a graph and its arboricity as:

\begin{lemma}[\cite{CliqueCountEden2020}]\label{lem:TalyaSheshlemma} Let $\alpha$ be $G$'s arboricity, and $c_k$ a clique of size $k$, we have:
    \[\degreetalya(\set{c_k})\leq 2m\cdot  \alpha^{k-1}\enspace, \]
    where $\degreetalya(\set{c_k})=\sum_{g\in \set{c_k}}\degreetalya(g)$.
\end{lemma}

Using the extended definition of degree with respect to a segmentation, %In this paper we extend the prior definition for degree of a clique to degree of a motif w.r.t. a segmentation%, denoted by $\degree_{\seg_i}$,(see \cref{def:degree}), 
we extend the above result for dense motifs (the proof will be presented in the extended version):

\begin{lemma}\label{lem:arboricity} Let $\alpha$ be $G$'s arboricity and $\g$ be a motif of size $k$ with segmentation $\seg(\g)=\langle \seg_2,\seg_3,\dots , \seg_{k-1}, \seg_k (=\g)\rangle $ (see \cref{Def:segment}) that is \csmooth, i.e., the minimum degree of $\seg_i$ is $\delta(\seg_i) = i - 1 - \eem$. Then for an arbitrary $2\leq i\leq k$, 
    %\[\degreei{i}{\seg}(\set{\seg_i})\leq 2m\cdot \left((\eem+1)\cdot  \alpha\right)^{i + 1 - \eem)}\cdot n^{\eem}\enspace, \]
    \[\degreei(\set{\seg_i})\leq 2m\cdot \left((\eem+1)\cdot  \alpha\right)^{i - 1 - \eem}\cdot n^{\eem}\enspace, \]
where $\degreei(\set{\seg_i})=\sum_{g\in \set{\seg_i}}\degreei(g)$.
\end{lemma}

 Note that for a clique, a segmentation is  a sequence of smaller cliques with $\eem = 0$ so our result is a generalization of \cref{lem:TalyaSheshlemma}.

%in general, our task is to find a collection $R\subseteq E$ such that: \emph{goal 1.} average of $f$ in  $R$ can be estimated using a sub-linear number of queries, \emph{goal 2.} while size of $R$ is sub-linear, $\bar{f}(R)$ is close to $\bar{f}(E)$. In section \ref{algo}, by using the assignment rule and leveraging the \emph{degeneracy} of a graph we achieve goal 1.{ \color{blue} TALK ABOUT VARIANCE} Here we show how to use MCMC to find $R$ with the desired property (goal 2). 

%Fixing a motif $g_k$, with $k$ vertices, for each $e$ we define $f(e)$ to be the number of instances of $g_k$ in $G$ that are \emph{assigned} to $e$.

%We assign any instance of $g_k$ in $G$ inductively to its sub-motifs until we reach an edge in $E$ through an \emph{assignment rule} so that at the end each $g_k$ is  assigned to one edge and only one edge in $E$. Our assignment rule is presented in \cref{Def:assign} and it is based on a $k$ segmentation of $g_k$ (see \cref{Def:segment}).

%\section{Motif segmentation and hierarchical assignment rule}\label{sec:hir}
\section{\algoname}\label{sec:algo}

%Thus, 

%\subsection{\algonamelayers: Hierarchical sampling of  $\mathbf{\g}$ based on ${\mathbf\seg(\g)}$. } 

%Assume from here on that $\g$ is given and $\seg(\g)$ is pre-determined. 
In this section, we present \algoname, our suggested algorithm for  \textsc{D}ense \textsc{M}otif count \textsc{E}s\textsc{T}imation through \textsc{R}andom \textsc{I}ncident \textsc{S}ampling, and we analyse its correctness and efficiency. A key component is the hierarchical sampling procedure 
%whose pseudocode is 
presented in \cref{alg:layers}. 

%\pragraph{\textbf{ Overview of \algonamelayers. }}
Given a $\csmooth$ motif $\g$ along with its segmentation $\seg(\g)$, an initial set of edges ${\mathcal E}\subseteq E$ and numbers $l_3,\dots, l_k$, $\algonamelayers$ \emph{extends} the edges to higher order sub-graphs until reaching sub-graphs isomorphic to $\g$. In this process, it constructs layers $\layer{2},\layer{3},\dots ,\layer{k}$  such that each $\layer{i}$ is a collections of sub-graphs isomorphic to $\seg_{i}(\g)$.
The construction of layers is an inductive procedure: we let $\layer{2}={\mathcal E}$, and calculate $\degreei(\layer{2})\doteq \sum_{g\in \layer{2}}\degreei(g)$  (lines 1-2 of \cref{alg:layers}). 

For $i=2,3,4,\dots, k-1$, we construct $\layer{i+1}$ by sampling $g\in \layer{i}$ from distribution $\p_{i}$ which is defined to satisfy $\p_{i} (g)\propto \degreei(g)$. Using $\Or{G}$ we sample $u$ u.a.r. from $ \Nseg(g)$. We let $g'=g + u$, and check if $g'\equiv \seg_{i}$ and $g=\assigi{i}(g')$. If both of these conditions are satisfied, $g'$ will be added to collection $\layer{i+1}$ (lines 7-14 in \cref{alg:layers}).

The output of $\algonamelayers$ is $\layer{k}$ which is a collection of sub-graphs in $\set{\g}$. Furthermore each $g\in \layer{k}$ is assigned uniquely through $\assig=\assigi{2}\circ \assigi{3}\circ\dots \circ \assigi{k-1} $ to an edge  $e\in {\mathcal E}$.

In order to prove the correctness of our method, we  ensure that all layers are ``\emph{nice}'' which means they are large enough to carry enough information to the next layer (see \cref{sec:analysis}). 
%We prove the niceness of each layer  in \cref{sec:analysis}.

\begin{remark}[\textsc{EstimateEdgeCount}]
The number of edges of the graph can be approximated, w.p. $1-o(1)$ and within error $\epsilon$, in the random walk model by running $\frac{\log n}{\epsilon^2} \sqrt{m}$ independent random walks and estimating $m$ by counting collisions; see e.g., \cite{triangleBerakdd,HamouPeresSODA}.  

%and using the birthday paradox
%Some other works \cite{HamouPeresSODA} suggested a method based on restarting random walks with , roughly, the same complexity.
%. Most of available methods have 
%The query complexity of these works is $\frac{\log n}{\epsilon^2} \sqrt{m}\tmix$~. 
\end{remark}

\begin{algorithm}
\begin{algorithmic}[1]  
\REQUIRE ${\mathcal E},\seg, l_3, l_4\dots, l_k$
% \ENSURE $g_{3_{1}}, g_{3_{2}}$
\STATE ${\layer{2}}=\mathcal{E}$
\STATE Calculate ${\degree}({\layer{2}})$ as ${\degree}({\layer{2}})=\sum_{e\in {\mathcal E}}\degree(e)$ 
\FOR{$i = {3}$ to $ k$}
    \STATE $\layer{i}\gets \emptyset$, $\degreei(\layer{i}) \gets 0$
    \STATE // \COMMENT  {\small Construct $\layer{i}$ by sampling  $g\in\layer{i-1}$  and extending to  $g'\in 
    \set{\seg_{i}}$.} 
    \FOR{$j=1$ to $l_{i}$}
        \STATE Sample $g$ from $\layer{i-1}$ w.p. $\degreei(g)/{\degreei(\layer{i-1})}$. \label{line:alg-g}
        \STATE For each  vertex $v$ in $g$ invoke $\Or{G}(v)$  to obtain $\Nseg(g)$.
        \STATE Sample $u\sim \Nseg(g)$ u.a.r.
        \STATE $g' \gets g + u $. \label{line:alg-gp}
        \IF {$g'\in \set{\sigma_i}$ and $g=\assigi{i}(g')  $} \label{line:alg-if}
                    \STATE $Y_{i, j}=1$
            \STATE $\layer{i}=\layer{i} \cup g'$
            \STATE ${\degreei}(\layer{i})= {\degreei}(\layer{i})+\degreei(g')$. 
        \ENDIF
    \ENDFOR
    \STATE $Y_i = \sum_{j=1}^{l_i} Y_{i,j}$
\ENDFOR
\RETURN{$\langle \layer{k}, Y_k, \degreei(\layer{2}), \degreei(\layer{3}),\dots \degreei(\layer{k})\rangle $}
\end{algorithmic}
\caption{$\algonamelayers(\mathcal{E},\seg, l_3,\dots, l_k)$
\label{alg:layers}
}
\end{algorithm}

\begin{algorithm}
\begin{algorithmic}[1]
\REQUIRE $G=\langle V,E\rangle , \g, \seg(\g),  r, \tmix, \varepsilon, l_3, \cdots, l_k$
\ENSURE $s$
\STATE $s\in V \leftarrow$ an arbitrary start point for random walk.
\STATE $\tilde{r}\leftarrow r\cdot \tmix$
\STATE $\mathcal{E} \leftarrow$ multi-set of edges on random $\tilde{r}$-walk from $s$. 
%\textcolor{red}
\STATE $\langle \layer{k},Y,{\degreei}_{{2}}, \dots , {\degreei}_{k} \rangle  = \textsc{ConstructLayers}(\mathcal{E},\seg, l_3,l_4,\dots ,l_k)$
\STATE $\widehat{m} = $ \textsc{EstimateEdgeCount}$(\mathcal{E}, \tmix)$

    \STATE $\estimate = Y\cdot  \frac{\widehat{m}}{\tilde{r}} \label{line:alg-output} \nicefrac{\prod_{i=3}^{k}{\degreei}_{{i-1}}}{\prod_{i=3}^{k}l_i} $
\RETURN{$\estimate$}
\end{algorithmic}
\caption{\algoname\label{alg:count}}
\end{algorithm}

\iffalse 
\shahrzad{The rest of this section needs more work}
\shahrzad{
Defining $f^{(i)}_{\assig}:\set{\seg_i}\rightarrow \set{\g}$ as $f^{(i)}_{\assig}(g)\doteq \abs{\{g\in \set{\seg_{i}}; g'=\assigi{i}\circ\dots \circ \assigi{k-1}(g) \}}$, we can re-define $f_{\assig}:E\rightarrow \set{\g}$ as: $f(e)=\prod_{i=2}^{k-1}f^{i}$
?????
, we will have that $\bar{f}$ is an unbiased ??}

\shahrzad{
After constructing these layers, \algoname~  estimates number of $g_k$'s copies by sampling  them from last layer. We bound the query complexity of \algoname~ knowing that ${\mathcal D}_j({\mathcal G}_k)$ is small for all $j$s in a $c$-smooth segmentations. 
This is shown in following lemma (proved in the appendix),  which extends the seminal result  of Chiba and Nishizeki \cite{ChibaN85} to graphlets amenable to a $c$-smooth segmentation
(presented in detail \cref{sec:algo}).}

\shahrzad{This lemma may be wrong}
\begin{lemma}\label{lem:chibAext}
Consider graphlet $g_k$  and assume there exists a $c$-smooth segmentation  for it namely $seg=\langle seg_2,seg_3,\dots ,seg_{k-1} \rangle$. For any $2\leq j\leq k-1$ we have: 
\begin{equation}
{\mathcal D}_j({\mathcal G}_k)=\sum_{g \in {\mathcal G}_k}d_j(g)\leq 2m^c \alpha^{k-c}
\end{equation}
, where $m$  and $\alpha$ are respectively the number of edges in $G$ and its degeneracy.
\end{lemma}

\fi

\subsection{Theoretical Analysis of \algoname}\label{sec:analysis}

\begin{table}
\small
\centering
\begin{tabular}{ | c | p{7cm} |}
\toprule
{\bf Symbol} & \hspace{2.5cm}{\bf Definition} \\ 
\midrule
$\seg(\g)/\seg$ & \csmooth~segmentation of $\g$, $\seg = \seg_2, \cdots, \seg_k (= \g)$.\\
$\set{\g}$ & Set of occurrences of an arbitrary motif $\g$. \\
$T$ & Size of $\set{\g}$, i.e., $|\set{\g}|$.  \\ 
$T_i$ & Size of $\set{\seg_i}$, i.e., $|\set{\seg_i}|$.  \\ 
$d(S)$ & Size of neighborhood of $S$, i.e., $|\Ne(S)|$. \\
$\rep(g)$ & Representative of $g$ w.r.t. $\seg$; $c+1$ nodes in $g$ w\ min neighbors. \\ 
$\degreei(g)$ & Degree of representative of $g$ w.r.t. $\seg$, i.e., $\degree(\rep(g))$.\\
$\degreei_{i}$ & Degree of layer $\layer{i}$ returned by the algorithm.\\
%$\assig(g')$ & A collection of $c+1$ vertices in $g$ with minimum neighborhood degree. & ??\\ 
$\assigi{i}(g)$ & Assignment function mapping $\set{\seg_i}$ to %instance in  
$\set{\seg_{i-1}}$. \\ 
$\f{i}(g_i)$ & Number of copies of $\set{\g}$ that are assigned to $g_i\in\set{\seg_i}$. \\ 
$\fmax{i}$ & Max number of copies of $\set{\g}$ assigned to a graph $\set{\seg_i}$. \\
$F_{\max}$ & $\max_i \fmax{i}$. \\
\bottomrule
\end{tabular}
\caption{\small Table of definitions}\label{table:def}
\end{table}

\iffalse
\begin{table*}
\small
\centering
\begin{tabular}{ | c | c |}
\toprule
Symbol & Definition \\ 
\midrule
$\seg(\g) = \seg_2, \seg_3, \cdots, \seg_k (= \g)$ & \csmooth segmentation of $\g$\\
$\set{\g}$ & set of occurrences of an arbitrary motif $\g$ \\
$T$ & size of $\set{\g}$, i.e., $|\set{\g}|$  \\ 
$T_i$ & size of $\set{\seg_i}$, i.e., $|\set{\seg_i}|$  \\ 
$d(S)$ & size of neighborhood of $S$, i.e., $|\Ne(S)|$ \\
$\rep(g)$ & representative of $g$ w.r.t. $\seg$; $c+1$ vertices in $g$ with minimal neighbors. \\ 
$\degreei(g)$ & degree of representative of $g$ w.r.t. $\seg$, i.e., $\degree(\rep(g))$\\
%$\assig(g')$ & A collection of $c+1$ vertices in $g$ with minimum neighborhood degree. & ??\\ 
$\assigi{i}(g)$ & assignment function from $\set{\seg_i}$ to instance in $\set{\seg_{i-1}}$ \\ 
$\f{i}(g_i)$ & Number of copies of $\set{\g}$ that are assigned to $g_i\in\set{\seg_i}$ \\ 
$\fmax{i}$ & Max number of copies of $\set{\g}$ that are assigned to a graph $\set{\seg_i}$ \\
\bottomrule
\end{tabular}
\caption{Table of definitions}\label{table:def}
\end{table*}
\fi
In this section, we prove \Cref{thm:intro}.  Let $Y$ be the second output parameter of $\algonamelayers$ called in line 4 of \Cref{alg:count}. For any $i \geq 3$ we define $c_i\doteq \frac{m}{\tilde{r}} \cdot  \prod_{j=3}^{i-1} \frac{\degreei_{j-1}}{l_j} \cdot \degree_{i-1}$.  Therefore, the output of \algoname\ is $\widehat{T}=Y\cdot \frac{c_k}{l_k}$~. We show in \Cref{lem:unbiased}, whose proof will be presented in an extended version of the paper, that this is an unbiased estimator. 

\begin{lemma}\label{lem:unbiased}
Let $\estimate$ be the outcome of \algoname, then $\ex[\estimate]=T \enspace.$ %\[\ex[\estimate]=T \enspace.\]
\end{lemma}

To show that $\widehat{T}$ is concentrated around its mean with high probability we first show  that with high probability we have
\begin{equation}\label{eq:Y}Y\in(1\pm \epsilon)l_{k}\cdot \frac{\f{k-1}(\layer{k-1})}{\degree_{k-1}}~.\end{equation}
Here $\f{k-1}(\layer{k-1})=\sum_{g\in \layer{k-1}} \f{k-1}(g)$, thus it is the total number of copies of $\g$ that are assigned to some sub-motif in layer $k-1$.
 The reader may wonder how \cref{eq:Y} helps in counting copies of $\g$. \cref{lemma:YY} shows that by maintaining  niceness properties  in the layers, which we define later (in \cref{eq:goodcond2}), from \Cref{eq:Y} we can conclude  that with high probability 
$Y\cdot \frac{c_k}{l_k}\simeq T$~.

\iffalse 
\begin{lemma}\label{lem:y}
Let $Y$ be the random variable  in line \ref{line:alg-output} of \algoname, then w.p. $1-o(1)$ \cref{eq:Y} holds 
%Conditioned on  $\bar{r}\geq $ and $\layer{3},\layer{4},\dots \layer{k-1}$ being nice,   we have,
%\[
%\cP{Y\in(1\pm \epsilon)^{i+1} l_{k}\cdot \left[ \frac{\epsilon}{\log(n)}\cdot \frac{T}{\degreei(\set{\seg_{k-1}})}\right]}
%\geq 1-o(1)\]
\iffalse 
\begin{equation}\label{eq:raneY}
\cP{Y\in(1\pm \epsilon)l_{k}\cdot \frac{\f{k-1}(\layer{k-1})}{\degree_{k-1}}}
\geq 1- o(1)    
\end{equation}
\fi 
%\[
%\cP{Y\in(1\pm \epsilon)l_{k}\cdot \frac{\f{k-1}(\layer{k-1})}{\degreei_{k-1}}}
%\geq 1-o(1)\]
if 
$l_k\geq \frac{(3 \log^2 n)}{ \epsilon^3(1-\epsilon)^k }\cdot \frac{\degree(\set{\seg_{k-1}})}{T}$
%
%\shahrzad{I think this value of $l_k$ is wrong}$l_k\geq 
%{(3 \log^2 n) } \cdot (\epsilon^2 )\cdot \frac{\degreei(\set{\seg_{k-1}}}{\abs{\set{\g}}}
%$ 
and for $3\leq i\leq k$,
$l_i\geq \frac{\log n^3}{\epsilon^3(1-\epsilon)^i}\cdot \fmax{i}(\layer{i}) \cdot \frac{\degreei(\set{\seg_i})}{T}  $, and $\tilde{r}\geq \tmix+ \frac{\log n^3}{\epsilon^3}\cdot  \trel\fmax{2}\cdot \frac{2m}{T} $ ~.
\end{lemma}
\fi

\begin{lemma}\label{lemma:YY}
If $~\f{k-1}(\layer{k-1})\in (1\pm \epsilon)^{k-1}\cdot \frac{T}{c_{k-1}}$ (see \Cref{eq:goodcond2}), and \Cref{eq:Y} holds, then \[\widehat{T}=\frac{c_k}{l_k}\cdot Y\in (1\pm \epsilon)^k T~.\]
\end{lemma}
\iffalse

\fi 

\Cref{lemma:YY} will be proven in the extend version of this paper. 
As a result of them, we conclude that it is sufficient to generate layers such that $\layer{k-1}$ is nice. We now proceed to prove \cref{eq:Y}

%Since $\layer{k-1}$ is produced by a hierarchical procedure, our proof is inductive. 

\iffalse 
\subsection{Expectation of Output Value}\label{sec:ex}
In this section, we show that our algorithm is an unbiased estimator, i.e., the expected value of the returned value by our algorithm is the desired value. We mainly use the following lemma to prove this bound.

\begin{lemma}\label{lem:prob} Given $g \in \set{\seg_i}$,  the probability of $g$ being sampled in some iteration of constructing $\layer{i}$ is $\nicefrac{1}{c_i}$ where $c_i\doteq \frac{m}{\tilde{r}} \cdot  \prod_{j=3}^{i-1} \frac{\degreei_{j-1}}{l_j} \cdot \degree_{i-1}$ for $i \geq 3$.% and $c_2 = \frac{m}{\tilde{r}}$.
%\[\cP{g\in \layer{i}}=\frac{1}{c_i}\enspace.\]
\end{lemma}

\begin{proof}[Proof of \cref{lem:unbiased}]
Let $A_k(g)$ be the event that an arbitrary motif $g\in \set{\g}$ is in $\layer{k}$ after a run of \algonamelayers. By \cref{lem:prob}, we have 
\begin{align*}
\ex[Y]=  \sum_{g\in \set{\g}}\cP{A_k(g)} = \frac{T}{c_k}. 
\end{align*}

%where $\frac{m}{\tilde{r}} \cdot l_k \cdot  \prod_{j=3}^{k} \frac{\degreei_j}{l_j}$.
Using linearity of expectation   we have 
$
    \ex[\estimate]= \ex[Y]\cdot {c_k}
    =T. 
$
\end{proof}
\fi 
\subsubsection{High Probability Concentration }\label{sec:conc}
In this section, 
%we %focus on showing show the concentration of the returned value around its mean with high probability 
%and 
we prove \cref{eq:Y}. 
%Our main idea is to establish requirements on layers so that the estimator enjoys required properties.
Intuitively, we would like all the layers to inductively acquire nice properties; thus we define two main properties for each layer $\layer{i}$: %that help us in achieving this: 
(i) value of $\f{i}$ should be concentrated around its mean, and (ii) each layer to be \emph{rich enough} to carry the information we need to the next layer (needed for inductive step). Formally, we phrase these requirements as follows.
%From \cref{lem:unbiased} we know that the output of \algoname\ is an unbiased estimator for $\abs{\set{\g}}$. Note that this out-put is constructed through hierarchical sampling while each $\f{i}(\layer{i})$ remains close to its mean. Note that in \Cref{lem:fi} we saw  $\ex\left[ \f{3}(\layer{3})\mid R\right]=l_3\cdot \frac{\f{2}(R)}{\degree(R)}$. A similar equation holds for other layers in fact in \cref{lem:fi} we show for any two consecutive layers $\layer{i-1}$ and $\layer{i}$ that  $\ex\left[ \f{i}(\layer{i})\mid \layer{i-1}\right]=l_i\cdot \frac{\f{i-1}(\layer{i-1})}{\degree(\layer{i-1})}$.

\begin{defin}[a nice layer]\label{def:nice}
For $i\in \{2,3,\dots k\}$, we say  $\layer{i}\sqsubseteq \set{\seg_i}$  is \emph{nice}  if 

\begin{equation}\label{eq:goodcond2}
 \f{i}(\layer{i})\in (1\pm \epsilon)^i \cdot  \frac{l_i}{c_i}\cdot T\enspace, \tag{Nice-Range}
\end{equation}
\begin{equation}\label{eq:goodcond1}
    \frac{\f{i}(\layer{i})}{\degreei_i}\geq (1-\epsilon)^i\cdot \frac{\epsilon}{\log n}\cdot \frac{T}{\degreei(\set{\seg_i})}\enspace \tag{Nice-Ratio}%\tag{$Prop-\f$}
\end{equation}
, where $c_i\doteq \frac{m}{\tilde{r}} \cdot  \prod_{j=3}^{i-1} \frac{\degreei_{j-1}}{l_j}\cdot \degree_{i-1}$.

\end{defin}

%Note that by Markov's inequality we know that if a set is large enough, the value of a non-negative random variable defined on it cannot significantly exceed its expected value. Thus it is not surprising that   by controlling size of each layer we can ensure it is nice. 

%This is proven in the following lemma: 

The proof of the following lemmas are presented in the appendix. The analysis of $R$ requires bounding the co-variance of consecutive samples (employing \cref{lem:cov}) as they are generated by running a random walk. The analysis of the other layers is similar to the work of \cite{CliqueCountEden2020} and tailored to $\csmooth$ motifs. 

\begin{lemma}\label{lemma:nice-R}
If $\tilde{r}\geq \tmix+\trel\cdot \frac{\log n}{\epsilon^2}\cdot\frac{\fmax{2}\cdot m}{T} $ then $\layer{2}=R $ is nice. 
\end{lemma}

\begin{lemma}\label{lemma:nice}
Let $3\leq i\leq k$. If $\layer{i-1}$ is nice and $l_i\geq \frac{\log n}{\epsilon^2}\cdot \fmax{i}(\layer{i}) \cdot \frac{\degreei_{i-1}}{\f{i-1}(\layer{i-1})}  $, then $\layer{i}$ is also nice~.
\end{lemma}

Before we can prove the concentration for random variable $Y$, we need the following lemma:
%one more lemma on the expected value of $Y$ for a given layer $\layer{k-1}$.

\begin{lemma}\label{lemma:lastlayer}
Given a layer $\layer{k-1}$ constructed in \algonamelayers\, we have 
\[
\ex\left[ Y\mid \layer{k-1}\right]=l_k \cdot \frac{\f{k-1}( \layer{k-1})}{\degree_{k-1}}~.
\]
\end{lemma}

Now we have all the required ingredients for proving \cref{eq:Y}.
\iffalse
\begin{lemma}
Let $Y$ be the random variable  in line 2 of \algoname.
Conditioned on  $\bar{r}\geq $ and $\layer{3},\layer{4},\dots \layer{k-1}$ being nice,   we have,
\[
\cP{Y\in(1\pm \epsilon)l_{k}\cdot \frac{\abs{\f{k-1}(\layer{k-1})}}{\degree(\layer{k-1})}}
\geq 1-1/n\]
if  $l_k\geq 
{(3 \log^2 n) } \cdot (\epsilon^2 )\cdot \frac{\degree(\set{\seg_{k-1}})}{\abs{\set{\g}}}
$ .
\end{lemma}
\fi
\begin{proof}[Proof of \cref{eq:Y}]
%Let $Y_{i,j}$ be a random variable corresponding to whether if condition in line \ref{line:alg-if} is satisfied. So for random variable $Y_i$ 
%using Chernoff bound for i.i.d. random variables we have
%the sampled graph $g \in L_{k-1}$ and $g'$ 
%From \cref{lemma:lastlayer} we know that assuming  $\layer{k}$ is fixed (has already been sampled) $\mathbb{E}\left[Y\right]=l_k\cdot\frac{\abs{\f{k-1}(\layer{k-1})}}{\degree(\layer{k-1})}$. 
Employing the Chernoff bound for i.i.d. random variables and using \cref{lemma:lastlayer} %on the expected value of $Y$
, we have:
\[
\cP{\abs{Y-\ex{\left[Y\right]}}\geq \epsilon \ex{\left[Y\right]}}\leq 2 \  \exp\left(- \frac{\epsilon^2}{3}\cdot  l_k\cdot\frac{\f{k-1}(\layer{k-1})}{\degree_{k-1}} \right)
\]
By the choice of $l_i$'s in the statement of the lemma and applying \cref{lemma:nice} and \cref{lemma:nice-R}, we have that all layers are nice. Thus by by applying \cref{eq:goodcond2}, we obtain 
\[
\cP{\abs{Y-\ex{\left[Y\right]}}\geq \epsilon \ex{\left[Y\right]}}\leq 2 \  \exp\left(- \frac{\epsilon^3(1-\epsilon)^{k-1}}{3\log n}\cdot  l_k\cdot \frac{T}{\degree(\set{\seg_{k-1}})} \right)
\]

Thus 
$
\cP{\abs{Y-\ex{\left[Y\right]}}\geq \epsilon \ex{\left[Y\right]}}\leq 2 \  \exp\left(-O(\log(n))\right) 
$ if $l_k\geq 
\frac{3 \log^2 n}{\epsilon^3 (1-\epsilon)^{k-1}} \cdot \frac{\degree(\set{\seg_{k-1}})}{T}$.\end{proof}
%, so the result in the lemma follows. 

\begin{proof}[Proof of \cref{thm:intro}]
Note that by employing  \cref{lem:arboricity} and the fact that $\g$ is $\csmooth$ we have:
for all $i$, $\degree(\set{\seg_i})\leq 2m \cdot  n^{c}\left((c+1)\alpha\right)^{i-(c+1)}$.
Note  all $l_i$s are taken large enough to ensure the layers are nice. In particular from the niceness of $\layer{k-1}$ and   \cref{eq:goodcond2} we conclude that 
$\f{k-1}(\layer{k-1})\in (1\pm \epsilon)^{k-1}\cdot \frac{T}{c_{k-1}}$. 
From \cref{lemma:YY} we conclude that $\widehat{T}=\frac{c_k}{l_k}\cdot Y\in (1\pm \epsilon)^k T$.
Plugging in the values in \cref{eq:goodcond1} the minimum required size for each $l_i$ will be:
\begin{align*}
    l_i\geq  \frac{\log n^3}{\epsilon^3(1-\epsilon)^i}\cdot \fmax{i}(\layer{i}) \cdot \frac{2m\cdot n^c\cdot  \left((c+1)\alpha\right)^{i-(c+1)}}{T} \\\text{, and ~~}\quad
l_k\geq \frac{(3 \log^2 n)}{ \epsilon^3(1-\epsilon)^k }\cdot \frac{2m\cdot n^c\cdot ((c+1)\alpha)^{k-(c+1)}}{T}~. 
\end{align*}
\end{proof}

\section{Experimental Evaluations}
\label{sec:exp}
\begin{figure}
    \centering
    \includegraphics[width=0.38\textwidth]{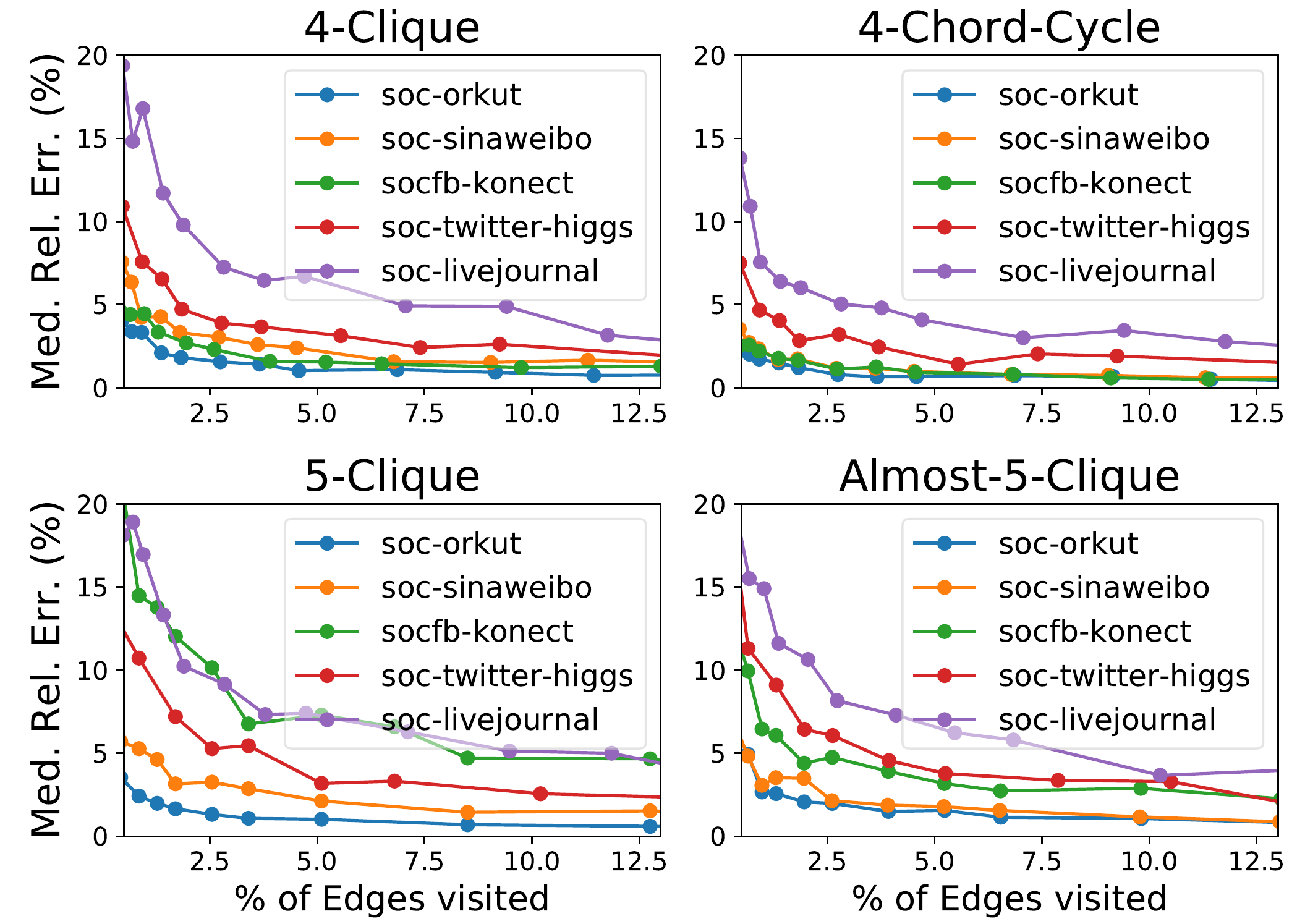}
    \caption{\small Plot of median relative error estimates for \algoname~ on various datasets and various motifs for 100 runs. We show the effect of varying the random walk length. We have median relative error percentage on y-axis, and percentage of the edges visited on x-axis.}
    \label{fig:median_relative_error_demet}
\end{figure}
\begin{figure*}
    \centering
    \includegraphics[width=0.90\textwidth]{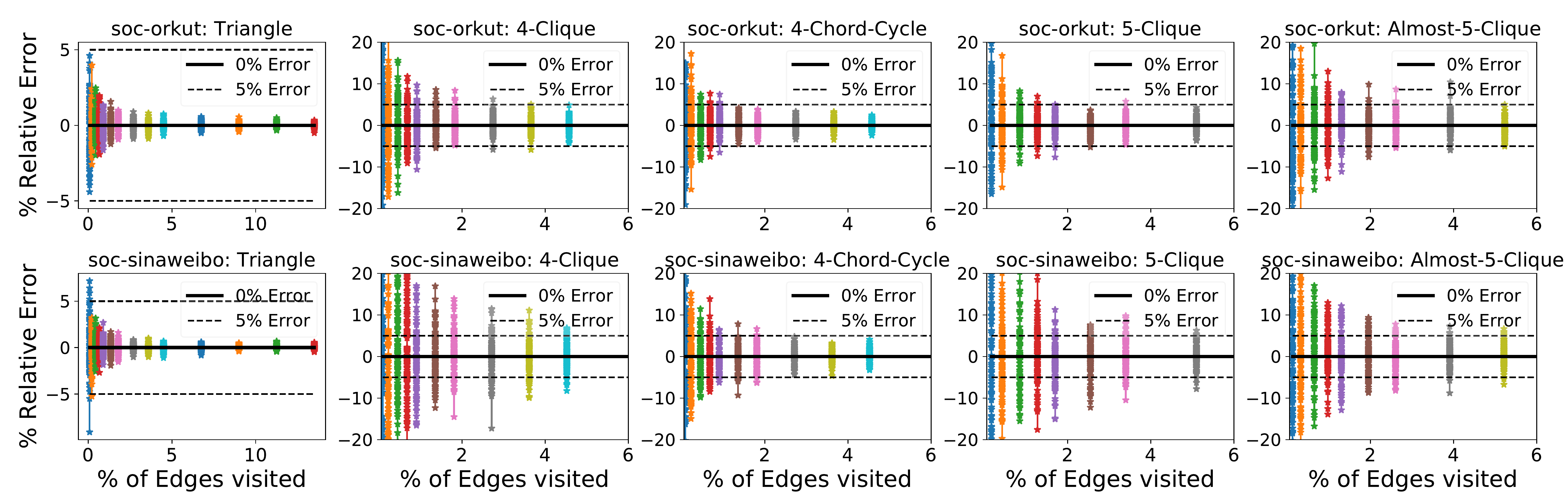}
    \caption{\small Convergence of \algoname~ for different motifs. On the y-axis, we plot relative error percentage of \algoname~ for each of the 100 runs corresponding to fixed length of random walk. On x-axis, we show the percentage of the queries made by \algoname~ during its execution by increasing the length of random walk. The maximum observed edge percentage corresponding to largest length of random walk is 6\%.}
    \label{fig:convergence_relative_error_demet}
\end{figure*}
\begin{figure*}
    \centering
    \includegraphics[width=0.85\textwidth]{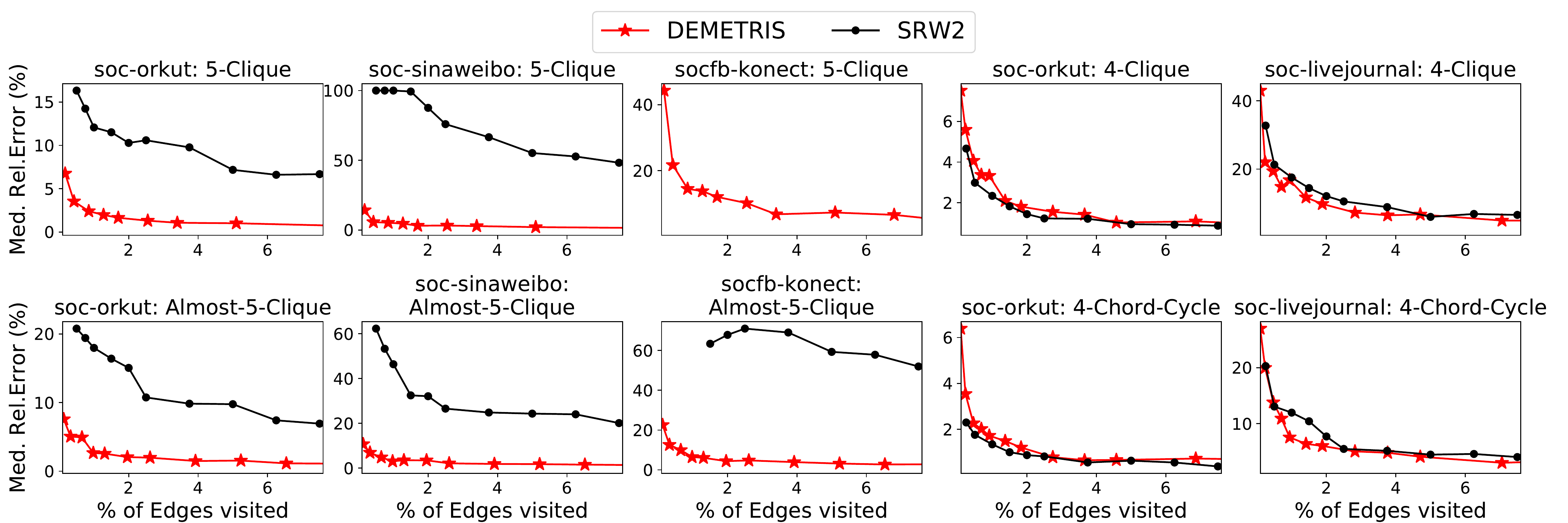}
    \caption{\small Comparison against baseline. For each dataset,  for each motif, and for a fixed length of random walk, we run both the algorithms 100 times. 
    We compare the median relative error in estimation vs the percentage of edges     visited.}
    \label{fig:comparison_demet_srw2}
\end{figure*}

In this section, we discuss the empirical performances of \algoname. 
Our implementation is in C++. We ran all the experiments on a workstation with 128GB DDR4 memory and Intel Xeon 2.20GHz processor running Ubuntu 20.
% \suman{@Jayesh: system specifications .}
We will make our code publicly available with the full version of the paper.
% \footnote{The anonymous code is available \href{https://www.dropbox.com/s/3yg4j2uyvp2jt8i/GraphletCountingUsingRW.zip?dl=0}{here}.}
% \footnote{The anonymous code is available \href{https://drive.google.com/drive/folders/1RANImoTwvYrJyAwwKGyjoRhug2wywYTq?usp=sharing}{here}.}
\footnote{The code repository is available \href{https://github.com/jayeshchoudhari/GraphletCountingUsingRW}{here}.}
We evaluate \algoname~ on the datasets given in~\Cref{table:dataset}.
We consider simple graphs by removing duplicate edges and self-loops.

\begin{table}[!ht]
  \caption{\small Description of our dataset with the key parameters, \#vertices($n$), \#edges($m$), \#triangles($g_{3,3}$), \#4-Cliques($g_{4,6}$), \#4-Chord Cycles($g_{4,5}$), \#5-Cliques($g_{5,10}$), and \#Almost 5-Cliques($g_{5,9}$).}
  \label{table:dataset}
  \begin{centering}
  \small
  \begin{tabular}{llllllll}
  \toprule
    {\bf Graph Name} & {\bf n}    & {\bf m}     & {\#$\mathbf{g_{3,3}}$} & {\#$\mathbf{g_{4,6}}$} & {\#$\mathbf{g_{4,5}}$} & {\#$\mathbf{g_{5,10}}$} & {\#$\mathbf{g_{5,9}}$} \\
    \toprule
    soc-orkut         & 3M   & 213M  & 525M        & 2.4B       & 33B      & 10.8B       & 84B      \\
    \hline
    \begin{tabular}[c]{@{}l@{}}soc-\\ sinaweibo\end{tabular}     & 59M  & 523M  & 213M        & 0.7B        & 27B      & 3.3B        & 47B      \\
    \hline
    \begin{tabular}[c]{@{}l@{}}socfb-\\ konect\end{tabular}      & 59M  & 185M  & 6.3M        & 0.5M        & 329M        & 36K        & 27M       \\
    \hline
    \begin{tabular}[c]{@{}l@{}}soc-\\ livejournal\end{tabular}   & 5M   & 85M   & 285M        & 9.9B       & 17B      & 467B         & 631B        \\
    \hline
    \begin{tabular}[c]{@{}l@{}}soc-twitter-\\ higgs\end{tabular} & 0.5M & 12.5M & 83M         & 0.4B        & 24B      & 2.2B        & 37.2B\\
    \bottomrule
\end{tabular}
\end{centering}
\end{table}
In our empirical evaluations, we seek to answer the following questions. 
\begin{itemize}[leftmargin=*]
    \item Is \algoname~ accurate for a suite of important motifs across multiple datasets? 
    
    We demonstrate in~\Cref{fig:median_relative_error_demet} that the
    accuracy of \algoname~ is remarkably high across multiple datasets for many different size cliques and 
    near-cliques. This is even more impressive considering the distributions of these cliques and near-cliques 
    vary significantly across datasets. For almost all the datasets, with less than $5\%$ of the edges, median error for \algoname~ is less than $5\%$.
    
    \item Does \algoname~converge to the true motif count given sufficient budget for the neighborhood samples? How much is the variance in the outcome? 
    
    We demonstrate the convergence of \algoname~ in~\Cref{fig:convergence_relative_error_demet}. As our theory suggests, \algoname~
    converges to the true count with increases in the number of queries. 
    
    \item Compared to existing algorithms for motif counting in the random walk model, how does \algoname~fare?\\
    We show in~\Cref{fig:comparison_demet_srw2} that \algoname~ consistently outperforms the baseline algorithms almost across all the datasets, and almost across all the different motifs. While observing only a tiny fraction of graph (around $2-3\%$), the accuracy achieved by \algoname~ is remarkable, where as the baseline performs at least twice worse than that of \algoname~ in most of the cases.
\end{itemize}

\mypar{Implementation Details}
\label{subsec:implementation}
The performance of \algoname~ is evaluated based on three different parameters: {\em Accuracy}, {\em Convergence}, and its {\em Comparison with the Baseline (SRW2)}.
To analyze the robustness of the estimation of  \algoname~  we evaluate its performance  in estimating the counts for different motifs: {\em Triangle (3-Clique)}, {\em 4-Clique}, {\em Almost 4-Clique (4-chord cycle)}, {\em 5-Clique}, {\em Almost 5-Clique}.
We use {\em relative error percentage}, defined as $\frac{(\textsc{ExactCount} - \textsc{AlgoEstimate})*100}{\textsc{ExactCount}}$, as a metric to analyze the performance of \algoname.
For analyzing the accuracy and the comparison with the baseline we use the median of the relative error percentage of 100 estimates, and to analyze the  convergence we represent the relative error percentage of all the 100 estimates.
For each of the 100 estimates the starting node of the random walk is chosen uniformly at random.
\algoname~ requires number of edges $m$ as a parameter along with the length of the random walk.
Note that, $m$ can be estimated from the random walk edges as similar to that in \cite{triangleBerakdd}.
In addition to the random walk length, the baseline SRW2 \cite{Chen2016} requires the sum of degree of edges $\sum_{e=(u,v) \in E}d_u + d_v$ as a parameter. We compute this separately and pass it as a parameter to SRW2.

\mypar{Accuracy} \algoname~ is eminently accurate across all the graphs with just at most $5\%$ of the edges of the graphs.
In Figure \ref{fig:median_relative_error_demet} we plot the median relative error percentage of \algoname~ over 100 runs for different motifs for each dataset while increasing the length of the random walk.
With the increase in the length of the random walk or the percentage of queries to around $5\%$, we observe that the median relative error percentage suddenly drops below $5\%$ for almost all the graphs and all different motifs. 

\mypar{{Convergence}} In Figure \ref{fig:convergence_relative_error_demet} we demonstrate the convergence of \algoname~ for \gName{orkut}, and \gName{sinaweibo} for different motifs.
We plot the relative error percentage of \algoname~ for 100 runs for a fixed length of the random walk.
 As we increase the percentage of edges visited, the relative error spread for \algoname~ is around or less than $5\%$ for almost all the motifs when we explore around $5\%$ of edges and tightly concentrates towards the $0\%$ relative error line which indicates the concentration towards the exact count. 

\mypar{{Comparison with Baseline}} We compare the performance of \algoname~ with SRW2 \cite{Chen2016}. 
Chen et al. \cite{Chen2016} propose a general framework to estimate statistics of graphlet of any size. 
This framework is based on collecting samples through consecutive steps of random walks.
SRW2 also requires the sum of the degree of edges ($\sum_{e=(u,v) \in E}d_u + d_v$) as  a parameter, and is computed separately.
% \jayesh{@Suman, should we add that for SRW2 we need sum of degrees? and also that for demet we use exact m?} \suman{Yes please do so. Discuss in the implementation details as well. Mention that $m$ can be computed from the collected random walk edges similar to the kdd triangle counting paper}

In Figure \ref{fig:comparison_demet_srw2} we compare the performance of \algoname~ and SRW2 with respect to the median relative error percentage.
Here we show the results for only a subset of graphs and motifs, for other graphs and motifs the performances are consistent and we defer the plots to 
%\cref{app:fig-x} and  
the full version of the paper. 
\algoname~ is consistently accurate across all the datasets and all different motifs.
We observe that in estimating the count of 5-Cliques for \gName{socfb-konect} dataset the estimation error for SRW2 is high.
And this is because the number of 5-Cliques present in the \gName{socfb-konect} dataset is very low ($36K$), which affects the performance of SRW2 even after exploring the around $5-7\%$ of the graph.
SRW2 performs equivalent to or slightly better than \algoname~ on the \gName{soc-livejournal} dataset across all the motifs and on the \gName{soc-orkut} dataset for {\em 4-Chord cycle} and {\em 4-Clique}.
In all the other cases, \algoname~ significantly outperforms SRW2 in terms of the median relative percentage error.
Across all the graphs and all the different motifs, after exploring only around $2\%$ of the graph, the median relative error percentage for \algoname~ is around $5\%$ or less. 
Whereas, for SRW2, for most of the cases, the median relative percentage error is more than $10\%$.

\iffalse
\begin{figure*}
    \centering
    \includegraphics[width=0.99\textwidth]{images/demet_median_relative_error.pdf}
    \caption{We plot median relative error estimates for \algoname~ on various datasets and various motifs for 100 runs. We show the effect of varying the random walk length. On y-axis, we have median relative error percentage, and on x-axis we have percentage of the edges visited.}
    \label{fig:median_relative_error_demet}
\end{figure*}
\fi

\section{Conclusion}

We study the problem of counting (near) cliques in a novel practical setting known as random walk  model. We build upon  prior work on theoretical study of sub-linear  algorithms  and develop new theory to address new challenges arising in our more realistic setting. We believe that the study of algorithms in the random walk setting is essential in the development of robust and practical algorithms.
%, and that our work is a contribution towards exploring a promising and exciting new area of research.

\section{Acknowledgments}
This work was carried out when Haddadan was a postdoctoral researcher at Brown University and she was supported by the Data Science Initiative at Brown University and by the NSF-TRIPODS grant.
%\newpage
\bibliographystyle{ACM-Reference-Format}
\bibliography{refs}

\appendix
%\subsection{Proof of Lemma \ref{lem:arboricity}}
\section{Missing Proofs}

In this section we present all missing proofs from the main sections. 

\subsection{Missing proofs from Section 2.1 }

\Cref{lem:arboricity} presented an upper bound for the number of occurrences of a $\rm c$-dense motif in a graph $G$ w.r.t.  $G$'s arboricity and $c$ thus extending a result of \cite{CliqueCountEden2020}. Here we present a proof: 
\begin{proof}[Proof of \Cref{lem:arboricity}]
Let $\prec_{\rm arb}$ be the acyclic ordering on vertices of $G$ where the cout-degree of each vertex is $\alpha$ (this ordering exists by the definition of arboricity \cite{matula1983smallest}).  For any $g\in \set{\seg_j}$, let $S(g)$ be the first $\eem+1$ vertices in $g$ w.r.t. $\prec_{\rm arb}$. Note that since each vertex in $g$ is not connected to at most $\eem$ other vertices, any  vertex in $g\setminus S(g)$ is connected to at least one vertex in $S(g)$. Moreover, let $\Ne_{\prec_{\rm arb}}(S)$ be the neighborhood of $S$ by outgoing edges, then since each vertex in $S$ has out-degree at most $\alpha$, $|\Ne_{\prec_{\rm arb}}(S)| \leq (c+1) \alpha$. For a set $S\subset V$ with $\eem+1$ vertices, define $\Gamma_{\seg_j}(S) = \{g\in \set{\seg_j}: S(g) = S\}$, then we have 
% \[
$\abs{\Gamma_{\seg_j}(S)}\leq \abs{\Ne_{\prec_{\rm arb}}(S)}^{j-(\eem+1)}\leq \left((\eem +1)\cdot \alpha\right)^{j-(\eem+1)}\enspace. $
% \]
Where the first inequality follows from the fact that $S$  needs to add $j-(c+1)$ to construct $\sigma_j$ and for each vertex there are $|\Ne_{\prec_{\rm arb}}(S)|$ possible choices. 
 
 Therefore we have:

 \begin{align*}
   \degree\left(\set{\seg_i} \right)  &=\sum_{g\in \set{\seg_i}}\degree(g)=\sum_{g\in \set{\seg_i}}\degree(\rep(g))
    \leq \sum_{g\in \set{\seg_i}}\degree(S(g)) \\
 %  =&\sum_{
 %  \substack{S\subset V;\\\abs{S}=\eem+1}} \sum_{g\in \mathcal{G}_S(\g)}\degree_{\seg_j}(g)\\
  % = & \sum_{
  % \substack{S\subset V;\\\abs{S}=\eem+1}} \sum_{g\in \mathcal{G}_S(\g)}\abs{\Nseg(g)}\\
     &= \sum_{
   \substack{S\subset V;\\\abs{S}=\eem+1}} \sum_{g\in \Gamma_{\seg_i}(S)}\abs{\Ne(S)}\\
 %   =   \sum_{ \substack{S\subset V;\\\abs{S}=\eem+1}}\abs{\Ne(S)}\cdot\abs{\mathcal{G}_S(\seg_i)}\\
   & \leq   \sum_{
   \substack{S\subset V;\\\abs{S}=\eem+1}} \abs{\Ne(S)}\cdot\left((\eem +1) \alpha\right)^{i-(\eem+1)}\\
    & \leq   \left((\eem +1) \alpha\right)^{i-(\eem+1)} \sum_{
   \substack{S;\\\abs{S}=\eem+1}} \sum_{v\in S}\dg(v)\\
    &\leq   \left((\eem +1) \alpha\right)^{i-(\eem+1)} \sum_{v\in V} \dg(v) \cdot \abs{\{S \subset V:\abs{S}=\eem+1, v\in S\}}\\
    &\leq   \left((\eem +1) \alpha\right)^{i-(\eem+1)} \sum_{v\in V} \dg(v) \cdot n^{\eem}  \\
    &\leq  \left((\eem +1) \alpha\right)^{i-(\eem+1)}\cdot  2m \cdot   n^{\eem} \qedhere
%\vspace{-1cm}
\end{align*}
\end{proof}
\subsection{Missing proofs from Section 3.1}
In this section, we present all the missing proofs that lead to the proof of \Cref{thm:intro}. 
  In the first subsection, we present all the proofs which lead to the proof of    \Cref{lem:unbiased}, which states that our estimator is unbiased. In the second subsection we show that our estimator is concentrated around its mean, in this proof we need to use that all layers are ``nice''. In the last subsection we show using an inductive proof that all the layers are nice. 

\subsubsection{Unbiased Estimator}
    \Cref{lem:unbiased} shows that  the output of \algoname, $\widehat{T}$, is an unbiased estimator for $T$. To prove it, we first show the following lemma:

\begin{lemma}\label{lem:prob}
Let
$g \in\set{\g}$, after running \algonamelayers, the probability that $g$ is sampled at any iteration of the final loop and added to $\layer{k}$ is equal to $1/c_k$~.
\end{lemma}
\begin{proof}
%[Proof of \cref{lem:prob}]
Let
$g \in \set{\g}$ and $g_2,g_3,\dots , g_k$ be such that for $i=2,3,\dots , k-1$ we have $g_i=\assigi{i+1}(g')$. Note that given $\seg(\g)$ and $\order$, for each $g \in \set{\g}$ the sequence of  $g_2,g_3,\dots , g_k$ will be determined uniquely, furthermore we have $g_i\in \set{\seg_i(\g)}$. Let 
$A_{i,\kappa}$ be the event that at iteration $\kappa$, $g_i$ is sampled and added to $\layer{i}$ and let $A_i$ be the total number of copies of $g_i$ in $ \layer{i}$. 
Using induction, we prove a more general statement that 
$\mathbb{P}(A_{i,\kappa})=1/c_i$ for any $\kappa=1,2,\dots, l_i$.

Assume the inductive hypothesis for $i-1$ and arbitrary $k'=1:l_{i-1}$. 

\begin{align*}
    \cP{A_{i,\kappa}} &= \left(\frac{\degreei(g_{i-1})}{\degreei_{i-1}}\cdot \frac{1}{\degreei({g_{j-1})}}\right) \cdot A_{i-1}\\
    &= \frac{1}{\degreei_{i-1}} \cdot \sum_{\kappa'=1}^{l_{i-1}} \cP{A_{i-1,\kappa'}}
    % &= \frac{l_{i-1}}{\degreei_i} \cdot \cP{A_{i-1,\kappa'}}\\
    = \frac{1}{\degreei_{i-1}} \cdot l_{i-1} \cdot \frac{\tilde{r}}{m} \cdot \prod_{j=3}^{i-2} \frac{l_j}{\degreei_{j-1}}\cdot \frac{1}{\degreei_{i-2}}\\
    & = \frac{\tilde{r}}{m} \cdot \frac{1}{\degree_{i-1}} \cdot \prod_{j=3}^{i-1} \frac{l_j}{\degreei_{j-1}} = \frac{1}{c_i} 
    %&=\sum_{\kappa'=1}^{l_{i-1}}\cdot \cP{A_{i-1,\kappa'}}\\
    %   &=\sum_{\kappa'=3}^{l_{i-1}}\left(  \frac{\degreei(g_{j-1})}{\degree(\layer{j-1})}\cdot \frac{1}{\degreei({g_{j-1})}}\right)\cdot \cP{A_{i-1,\kappa'}}\\           &=\sum_{\kappa'=3}^{l_{i-1}}\left(  \frac{1}{\degree(\layer{j-1})}\right)\cdot \cP{A_{i-1,\kappa'}}\\ 
\end{align*}

where the fourth equality follows from enforcing the induction hypothesis and the fact that the rounds are independent.  
%\sara{Need to argue for i = 3 the case holds.}
\iffalse
Using induction hypothesis on $A_{i-1,\kappa'}$ we have:
\begin{align*}
    \cP{A_{i,\kappa}}          
    &=\left(  \frac{l_{i-1}}{\degree(\layer{j-1})}\right)\cdot 
    (\frac{\tilde{r}}{m}) \left( \prod_{j=3}^{i-2} \frac{l_j}{\degree(\layer{j-1})}\right)\cdot \frac{1}{\degree(\layer{i-2})}\\
    &=
    (\frac{\tilde{r}}{m}) \left( \prod_{j=3}^{i-1} \frac{l_j}{\degree(\layer{j-1})}\right)\cdot \frac{1}{\degree(\layer{i-1})}\enspace. 
\end{align*}
\fi
\end{proof}

\begin{proof}[Proof of \Cref{lem:unbiased}]
Note that we have $\widehat{T}=Y\cdot \frac{c_k}{l_k}$, so it suffices to show $\mathbb{E}[Y]=T\frac{l_k}{c_k}$.
Consider the loop in which we collect the last layer. 
Let $Y(i,g)$ denote the random variable indicating whether some $g\in \set{\g}$ has been sampled at iteration $i$
\begin{align*}
    \mathbb{E}[Y]=\sum_{g\in \set{\g}}\sum_{\kappa=1}^{l_k}   \mathbb{E}[Y(i,g)]=\abs{\set{\g}}\sum_{\kappa=1}^{l_k} 1/c_k= T \cdot \frac{l_k}{c_k}
\end{align*}
\end{proof}

%\subsection{Proof of Lemma \ref{lemma:nice-R}} 

\subsubsection{High 
 Probability Concentration}
In this subsection we prove that if the last layer is nice then we have high probability concentration for $\widehat{T}$. In other words we prove  \cref{lemma:YY} which states: 

If $~\f{k-1}(\layer{k-1})\in (1\pm \epsilon)^{k-1}\cdot \frac{T}{c_{k-1}}$ (see \Cref{eq:goodcond2}), and \Cref{eq:Y} holds, then \[\widehat{T}=\frac{c_k}{l_k}\cdot Y\in (1\pm \epsilon)^k T~.\]

\begin{proof}[Proof of \cref{lemma:YY}]
Note that $c_k=c_{k-1}\cdot \frac{\degree_{k-1}}{l_{k-1}}$, thus,
\begin{align*}
\widehat{T}=Y\cdot \frac{c_k}{l_k}\in &\  (1\pm\epsilon )\cdot \frac{c_k}{l_k}\cdot \frac{l_k}{\degree(\layer{k-1})}\cdot \f{k-1}(\layer{k-1})\\
    \in &\   (1\pm\epsilon )\cdot (1\pm \epsilon)^{k-c_k}\cdot \frac{1}{\degree_{k-1}}\cdot \frac{l_{k-1}}{c_{k-1}}\cdot T
    \in  (1\pm \epsilon)^k T\enspace .
\end{align*}
\end{proof}

\subsubsection{Niceness of Layers}\label{sec:nice}
We now prove \Cref{lemma:nice} and \Cref{lemma:nice-R}, and a more general version of \Cref{lemma:lastlayer}.
In order to prove these lemmas we first establish bounds on the expected value of degree of different layers and the expected value of total motif count of different layers. Consider $\g$ and a segmentation of it 
$\seg=\seg_2,\seg_3,\dots ,\seg_k$. 

\begin{lemma}\label{lemma:lay2}
We have $\ex\left[\degree_2\right]=\frac{\tilde{r}}{m}\cdot \degree(E)$, 
and for $i=3,4,\dots , k$ $\ex\left[\degree_i\right]=
\frac{l_i}{c_i}\cdot \degree(\set{\seg_i})$.
\end{lemma}
\begin{proof}%proof of lemma lemm:f
Using linearity of expectation we conclude:  
\[
\ex\left[\degree(R)\right]=\sum_{e\in E} \cP{e\in R}\degree(e)=\frac{\bar{r}}{m}\sum_{e\in E}\degree(e)= \frac{\bar{r}}{m}\degree(E)\enspace.
\]

Similarly  employing \cref{lem:prob} we have:

\begin{align*}
    \ex\left[\degree(\layer{i})\right]&= \sum_{g_i\in \set{\seg_i}}\sum_{j=1}^{l_i}\cP{g_i \text{ is \ sampled}}\degree(g_i)\\
    &=\frac{1}{c_i}\sum_{j=1}^{l_i}\cdot \sum_{g_i\in \set{\seg_i}} \degree(g_i)\\
    &= \ \frac{l_i}{c_i}\cdot \degree\left(\set{\seg_i}\right)\enspace . 
\end{align*}
\end{proof}

The proof of \Cref{lemma:nice-R} is a direct application of the following lemma:
\begin{lemma}\label{lemm:f} We have 
$\ex\left[
\f{2}(\layer{2})\right]=\frac{\tilde{r}}{m}\cdot T$ and $\mathbb{V}\left[ \f{2}(R)\right]\leq \frac{\trel}{2}\cdot  \fmax{2} \cdot  \frac{\tilde{r}}{m}\cdot T$.
%Furthermore, after $R$ is sampled we have $\ex\left[ \f{3}(\layer{3})\mid \layer{2}\right]=l_3\cdot \frac{\f{2}(\layer{2})}{\degree_2}$~.
\end{lemma}

\begin{proof}
Let $\layer{2}=\{e_1,e_2,\dots ,e_{\bar{r}}\}$  we have 
\begin{align*}
   \mathbb{E}\left[\f{2}(\layer{2})\right]&= \sum_{e\in E }\sum_{g\in \assig^{-1}(g) }\mathbb{P}(e \text{ is sampled})\\
   &= \sum_{e\in E }\sum_{g\in \assig^{-1}(g) }\frac{\bar{r}}{m}\\
   &= \frac{\bar{r}}{m}\cdot T~.
\end{align*}

Using \cref{lem:cov} we bound the variance as
\begin{align*}
\mathbb{V}\left[\f{2}(\layer{2})\right]
&\leq \sum_{i,j=1}^{\bar{r}}\mathbb{C}\left[\f{2}(e_i),\f{2}(e_{j})\right]
\leq2 \sum_{i=1}^{\bar{r}}\sum_{j=i}^{\bar{r}}\mathbb{C}\left[\f{2}(e_i),\f{2}(e_{j})\right]\\
&\leq2 \sum_{i=1}^{\bar{r}}\sum_{j=i}^{\bar{r}}\lambda^{j}\mathbb{V}\left[\f{2}(e_{i})\right]
\leq\frac{1}{2} \sum_{i=1}^{\bar{r}}\frac{1}{1-\lambda}\mathbb{V}\left[\f{2}(e_{i})\right]\\
&\leq 2 \sum_{i=1}^{\bar{r}}\frac{1}{1-\lambda}\fmax{2}\cdot \mathbb{E}\left[\f{2}(e_{i})\right]
\leq 2\trel\cdot  \fmax{2} \cdot  \frac{\tilde{r}}{m}\cdot \abs{\set{\g}}
\end{align*}

\end{proof}

 Now having these lemmas for the expected degree of layers and also concentration of number of motifs assigned to the starting layer, it remains to bound the expected value of motif counts of higher level which can be achieved by the following inductive lemma on layers. %we need one last lemma connecting the motif counts of higher levels to the starting layer. 

The following lemma generalized and extends \Cref{lemma:lastlayer}:

\begin{lemma}\label{lem:fi} 
Assume we \algonamelayers\ has reached the $i-1$-th iteration thus $\layer{i-1}$ is sampled. For the next iteration we have:
  \begin{equation}\label{eq:lemfieq5}
      \ex\left[ \f{i}(\layer{i})\mid \layer{i-1}\right]=l_i\cdot \frac{\f{i-1}(\layer{i-1})}{\degree(\layer{i-1})}
        \end{equation}
        
        Furthermore we have \begin{equation}\label{eq:lemfieq6}
 \mathbb{V}\left[\f{i}(\layer{i})\mid \layer{i-1}\right]\leq \fmax{i} \mathbb{E}\left[\f{i}(\layer{i})\mid \layer{i-1}\right]~.
        \end{equation}
\end{lemma}
\begin{proof}
We first prove \cref{eq:lemfieq5}. For an iteration $1 \leq j \leq l_i$, let $g_j$ denote the graph picked in line \ref{line:alg-g} and let $g'_j$ be the graph picked in line \ref{line:alg-gp}. So we have 
Note that % by definition 
%\[   \ex\left[\f{i}(\layer{i})\mid \layer{i-1}\right]=\ex\left[{\sum_{j = 1}^{l_i} Y_{i,j} \cdot \f{i}(g'_j)} \right]~.\]
 %  For each $g'$ in the above sum, let $g\in \layer{i-1}$ be such that $g=\assigi{i-1}(g')$ and let $w=g'\setminus g$. 
   %Since $e$ is determined uniquely we have $\f{3}(\zeta)=\sum$
   \begin{align*}
   \ex&\left[\f{i}(\layer{i})\mid \layer{i-1}\right] \\
   %& =\ex\left[{\sum_{j = 1}^{l_i} Y_{i,j} \cdot \f{i}(g'_j)} \right] \\
 %& =\sum_{j = 1}^{l_i} \ex\left[Y_{i,j} \cdot \f{i}(g'_j)\right] \\
    & =\sum_{j = 1}^{l_i} \sum_{\substack{g\in \layer{i-1}, u \in \Nseg(g): \\ g=\assigi{i}(g+u)}} \cP{g = g_j} \cdot \cP{u = g_j \setminus g_j'} \cdot  \f{i}(g + u) \\
    & =\sum_{j = 1}^{l_i} \sum_{\substack{g\in \layer{i-1}, u \in \Nseg(g): \\ g=\assigi{i}(g+u)}} \frac{\degreei(g)}{\degreei_{i-1}} \cdot \frac{1}{\degreei(g)} \cdot  \f{i}(g + u) \\
    & =\sum_{j = 1}^{l_i} \sum_{\substack{g\in \layer{i-1}, u \in \Nseg(g): \\ g=\assigi{i}(g+u)}} \frac{\f{i}(g+u)}{\degree_{i-1}}
    =l_i\cdot \frac{\f{i-1}(\layer{i-1})}{\degree_{i-1}}\enspace.
\end{align*}
where the last inequality follows from the fact that $\f{i-1}(g)$ is aggregation of $\f{i}(g')$ over $g'$ which gets assigned to $g$ and can be obtained from $g$ by adding some vertex in $u \in \Nseg(g)$.

\iffalse
   The above sum can be written as 
   \begin{align*}
\sum_{\substack{g\in \layer{i-1} \ 
\\
 \  w\in \Nseg(g),g=\assigi{i}(g+w)}} \sum_{cou=1}^{l_i}&\cP{g \text{ is sampled in } cou }\cdot 
\cP{w \text{ is sampled}} \cdot \f{i}(w+g)\\
\quad =&\sum_{\substack{g\in \layer{i-1} \ ,  \  w\in \Nseg(g)
\\
g=\assigi{i}(g+w)}} l_{i}\cdot \frac{\degree(g)}{\degree({\layer{i-1}})}   \cdot \frac{1}{\degree(g)}\cdot \f{i}(w+g)\\
 \quad =\ &\frac{l_{i}}{\degree(\layer{i-1})}\cdot  \sum_{\substack{g\in \layer{i-1} \ ,  \  w\in \Nseg(g)
\\
g=\assigi{i}(g+w)}} {\f{i}(w+g)}\\
 \quad =\ &l_i\cdot \frac{\f{i-1}(\layer{i-1})}{\degree(\layer{i-1})}\enspace . 
\end{align*}

Note that the last equation holds because

\fi
To see \cref{eq:lemfieq6} note that samples form $\layer{i}$ are collected independently. Thus we have:

\begin{align*}
&\mathbb{V}\left[\f{i}(\layer{i})  \bigm| \layer{i-1}\right]  
\leq
\ex\left[\f{i}(\layer{i})^2 \bigm|\layer{i-1}\right]\\
& = \sum_{g \in \layer{i}} \mathbb{E}\left[ \f{i}(g)^2\mid \layer{i-1}\right]
\leq 
\fmax{i} \cdot %l_i\cdot
\sum_{g \in \layer{i}}
\ex\left[\f{i}(g)\bigm|\layer{i-1}\right]\\
&= \fmax{i} \cdot\ex\left[\f{i}(\layer{i})\bigm|\layer{i-1}\right]~.
\end{align*}
\end{proof}
Now we have the ingredients for proving the main lemma. 
\begin{proof}[Proof of \cref{lemma:nice}]
Assume $\layer{i}$ is sampled. We now use Chebyshev's inequality to show concentration of $\f{i}(\layer{i})$. 

\begin{align*}
    &\cP{\abs{\f{i}(\layer{i})-\ex\left[\f{i}(\layer{i})\right]}\geq \epsilon \cdot \ex\left[\f{i}(\layer{i})\right] }
    \leq \frac{\mathbb{V}\left[\f{i}(\layer{i})\right]}{\epsilon^2 \ex\left[\f{i}(\layer{i})\right]^2}\\
    &\leq 
    \frac{\fmax{i}\cdot \ex\left[\f{i}(\layer{i})\right]}{\epsilon^2\cdot \ex\left[\f{i}(\layer{i})\right]^2}
    \leq 
    \frac{\fmax{i}}{\epsilon^2\cdot \ex\left[\f{i}(\layer{i})\right]}
    \leq 
    \frac{\fmax{i}}{\epsilon^2\cdot l_i\cdot \frac{\f{i-1}(\layer{i-1})}{\degree_{i-1}}}
\end{align*}

Taking $l_i\geq  \frac{\log n}{\epsilon^2}\cdot \fmax{i}\cdot \frac{\degree_{i-1}}{\f{i-1}(\layer{i-1})}$ we bound the above probability by 
 $\frac{1}{\log n}$~. Plugging in the value of $\ex\left[\f{i}(\layer{i})\right]$ we have:
 \[
 \mathbb{P}\left(\f{i}(\layer{i})\in (1\pm\epsilon)\frac{l_i}{\degree_{i-1}}\cdot \f{i-1}(\layer{i-1})\right)\geq 1-\frac{1}{\log n}\enspace . 
 \]
 
 We now use the assumption that $\layer{i-1}$ is nice. Using \cref{eq:goodcond2} we have $\f{i-1}(\layer{i-1})\in (1\pm\epsilon)^{i-1}\cdot \frac{l_{i-1}}{c_{i-1}}\cdot T$. Thus, w.p. at least $1-\frac{1}{\log n}$:
 
 \begin{align*}
     \f{i}(\layer{i})\in (1\pm\epsilon)\frac{l_i}{\degree(\layer{i-1})}\cdot \f{i-1}(\layer{i-1})\\
     \implies  \f{i}(\layer{i})\in (1\pm\epsilon)\frac{l_i}{\degree(\layer{i-1})}\cdot (1\pm\epsilon)^{i-1}\cdot \frac{T}{c_{i-1}/l_{i-1}}\\
      \implies  \f{i}(\layer{i})\in  (1\pm\epsilon)^{i}\cdot \frac{l_i}{c_{i}}\cdot T\enspace .\\
 \end{align*}
Thus we conclude that \cref{eq:goodcond2} holds w.p. $1-\frac{1}{\log n}$. Now for proving \cref{eq:goodcond2}, we divide both sides by $D_i$, so we get
\[
\frac{\f{i}(\layer{i})}{\degreei_{i}} \in  (1\pm\epsilon)^{i}\cdot \frac{T}{\degreei_{i}\cdot c_{i}/l_i}
\]
Since by \cref{lemma:lay2}, $\ex\left[\degree_i\right] = \frac{l_i}{c_i}\cdot \degreei(\set{\seg_i})$, by Markov inequality, 
\[
\cP{\degree_i \geq \log(n) \cdot \frac{\degreei(\set{\seg_i)}}{c_i/l_i}} \leq \log(n)^{-1}
\]
, so we get that 
% \[
$\frac{\f{i}(\layer{i})}{\degreei_{i}} \geq  (1-\epsilon)^{i}\cdot \frac{T}{\log(n) \degreei(\set{\seg_i})}$
% \]
holds with probability at least $1 - \log(n)^{-1}$.
%Now in order to prove the second condition of niceness, we just divide both sides by $\degreei_{i}$, so we get 

%Dividing both sides by \cref{eq:apx8} we  conclude \cref{eq:goodcond1} holds with probability at least $1-\frac{2}{(\log n)}$.
\iffalse
Using Markov's inequality we have 

\[ \cP{\degree(\layer{i})\geq {\log(n)} \ex\left[\degree(\layer{i})\right]}\leq{\log n} \enspace ,
\]

Plugging in the value of  from  in the above equation w.p. at least $1-\frac{1}{\log n}$ we have ,

\begin{equation}\label{eq:apx8}
\degree(\layer{i})\leq \frac{ \degree(\set{\seg_i})}{c_i}\cdot(\log n)\enspace. 
\end{equation}
\fi
\end{proof}

\begin{proof}[Proof of \cref{lemma:lastlayer}]  For an iteration $1 \leq j \leq l_i$, let $g_j$ denote the motif picked in line \ref{line:alg-g} and let $g'_j$ be the motif picked in line \ref{line:alg-gp} in iteration $j$ of constructing $\layer{i}$. $Y_{i,j}$ is an indicator random variable where $Y_{i,j}=1$ iff $g_j=\assigi{i}(g'_j)$ and $Y_{i,j}=0$ otherwise. By linearity of expectation, we have:
\begin{align*}
    &\ex\left[Y\mid \layer{k-1}\right]=\sum_{j=1}^{l_k}\cP{Y_{i,j} = 1\bigm| \layer{k-1}}\\
        &=\sum_{j=1}^{l_k}\sum_{\substack{g\in \layer{i-1}, \\ u \in \Nseg(g)}} \cP{g = g_j \text{, } u = g'_j\setminus g_j \text{,  and  } g_j=\assigi{i}(g'_j)}\\
           &=\sum_{j=1}^{l_k}\sum_{\substack{g\in \layer{i-1}, \\ u \in \Nseg(g)}}\frac{\degree(g)}{\degree
           _{k-1}}\cdot \frac{1}{\degree(g)} \cdot \frac{\f{k-1}(g)}{\degree(g)}\enspace\\  
                    &=\sum_{j=1}^{l_k}\frac{1}{\degree
           _{k-1}}\cdot\sum_{\substack{g\in \layer{i-1}, u \in \Nseg(g)}} \frac{\f{k-1}(g)}{\degree(g)}\enspace\\
                    &=\sum_{j=1}^{l_k}\frac{1}{\degree
           _{k-1}}\cdot\sum_{g\in \layer{i-1}} \abs{\Nseg(g)}\cdot \frac{\f{k-1}(g)}{\degree(g)}\enspace 
       %    &=\sum_{j=1}^{l_k}\sum_\sum_{\substack{g\in \layer{i-1}, \\ u \in \Nseg(g)}}\frac{\f{k-1}(g)}{\degree   _{k-1}}\enspace\\
    =l_k\cdot \frac{\f{k-1}(\layer{k-1})}{\degree
           _{k-1}}\enspace.
\end{align*}
\end{proof}

\section{Additional Experiments}\label{app:fig-x}
\begin{figure}[b]
    \centering
    \includegraphics[width=0.49\textwidth]{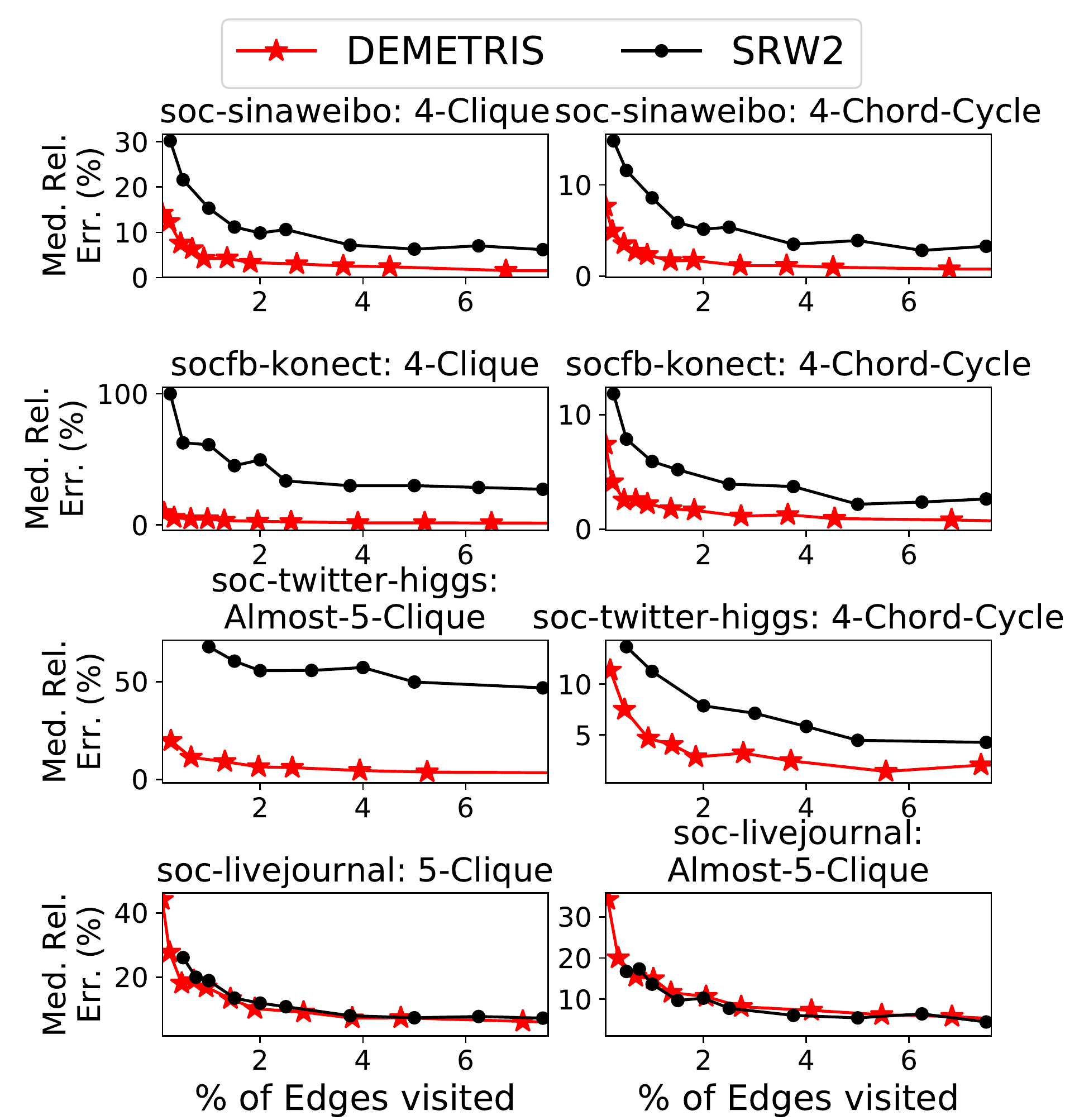}
    \caption{Comparison against baseline.}
    \label{fig:comparison_demet_srw2_2}
\end{figure}

Figure \ref{fig:comparison_demet_srw2_2} is the extension of Figure \cref{fig:comparison_demet_srw2} comparing the performances of \algoname~ and SRW2.
For each dataset, for each motif type, and for a fixed length of random walk, we run both the algorithms
100 times and then plot median relative error in estimation vs the percentage of edges visited.
As mentioned before, except for \gName{soc-livejournal} \algoname~ beats SRW2 significantly in terms of the median relative error percentage.
% \fi
\end{document}